\documentclass[aps,prc,10pt,tightenlines,notitlepage,showpacs,nofootinbib,superscriptaddress,twocolumn,eqsecnum,raggedbottom,preprintnumbers]{revtex4-1}

\usepackage{amsfonts}
\usepackage{amsmath}
\usepackage{amssymb}
\usepackage{array}
\usepackage{bbold}
\usepackage{bm}
\usepackage{booktabs}
\usepackage[dvipsnames,usenames]{color}
\usepackage{enumitem}
\usepackage{float}
\usepackage[margin=1in]{geometry}
\usepackage{graphicx}
\usepackage[utf8]{inputenc}
\usepackage{lastpage}
\usepackage{lipsum}
\usepackage{multirow}
\usepackage{nicefrac}
\usepackage{soul}
\usepackage{sidecap}
\usepackage[normalem]{ulem}
\usepackage[dvipsnames]{xcolor}

\usepackage{hyperref}
\hypersetup{
    colorlinks=true,
    urlcolor=blue,
    linkcolor = blue,
    urlcolor  = blue,
    citecolor = blue,
    anchorcolor = blue,
    pdftitle={electroproduction},
    pdfauthor={Maxim Mai, ...},
    pdfsubject={electroproduction}}

\renewcommand{\Re}{{\rm Re}}
\renewcommand{\Im}{{\rm Im}}

\makeatletter
\renewcommand{\p@subsection}{}
\renewcommand{\p@subsubsection}{}
\makeatother

\setlist[itemize]{leftmargin=10pt, itemsep=0pt}
\setlist[enumerate]{leftmargin=30pt, itemsep=0pt}
\setlist[description]{leftmargin=40pt}

\newcommand{\nocontentsline}[3]{}
\newcommand{\tocless}[2]{\bgroup\let\addcontentsline=\nocontentsline#1{#2}\egroup}

\makeatletter
\newcommand\appendix@section[1]{%
  \refstepcounter{section}%
  \orig@section*{{Appendix}:}%
}
\let\orig@section\section
\g@addto@macro\appendix{\let\section\appendix@section}
\makeatother

\newcolumntype{?}{!{\vrule width 2pt}}
\makeatletter
\newcommand{\thickhline}{%
    \noalign {\ifnum 0=`}\fi \hrule height 1pt
    \futurelet \reserved@a \@xhline
}
\newcolumntype{"}{@{\hskip\tabcolsep\vrule width 1pt\hskip\tabcolsep}}
\makeatother

\begin{document}

\date{\today}

\title{
J\"ulich-Bonn-Washington Model for Pion Electroproduction Multipoles
}

\author{M.~Mai}
\email{maximmai@gwu.edu}
\affiliation{Institute for Nuclear Studies and Department of Physics, The George Washington University, Washington, DC 20052, USA}

\author{M.~D\"oring}
\email{doring@gwu.edu}
\affiliation{Institute for Nuclear Studies and Department of Physics, The George Washington University, Washington, DC 20052, USA}
\affiliation{Thomas Jefferson National Accelerator Facility, Newport News, VA 23606, USA}

\author{C.~Granados}
\email{cggranadosj@gmail.com}
\affiliation{Institute for Nuclear Studies and Department of Physics, The George Washington University, Washington, DC 20052, USA}

\author{H.~Haberzettl}
\email{helmut@gwu.edu}
\affiliation{Institute for Nuclear Studies and Department of Physics, The George Washington University, Washington, DC 20052, USA}

\author{Ulf-G.~Mei{\ss}ner}
\email{meissner@hiskp.uni-bonn.de}
\affiliation{Helmholtz-Institut f\"ur Strahlen- und Kernphysik (Theorie) and Bethe Center for Theoretical Physics,  Universit\"at Bonn, 53115 Bonn, Germany}
\affiliation{Institute for Advanced Simulation and J\"ulich Center for Hadron Physics, Forschungszentrum J\"ulich,  52425 J\"ulich, Germany}
\affiliation{Tbilisi State University, 0186 Tbilisi, Georgia}

\author{D.~R\"onchen}
\email{d.roenchen@fz-juelich.de}
\affiliation{Institute for Advanced Simulation and J\"ulich Center for Hadron Physics, Forschungszentrum J\"ulich,
52425 J\"ulich, Germany}

\author{I.~Strakovsky}
\email{igor@gwu.edu}
\affiliation{Institute for Nuclear Studies and Department of Physics, The George Washington University, Washington, DC 20052, USA}

\author{R.~Workman}
\email{rworkman@gwu.edu}
\affiliation{Institute for Nuclear Studies and Department of Physics, The George Washington University, Washington, DC 20052, USA}

\collaboration{J\"ulich-Bonn-Washington}
\preprint{JLAB-THY-21-3348}

\begin{abstract}
Pion electroproduction off the proton is analyzed in a new framework based on a general parametrization
of transition amplitudes, including constraints from gauge invariance and threshold behavior. Data with energies
$1.13~{\rm GeV}<W<1.6~{\rm GeV}$ and $Q^2$ below $6~{\rm GeV}^2$ are included. The model is an extension of the latest J\"ulich-Bonn solution incorporating constraints from pion-induced and photoproduction data. Performing large scale fits ($\sim10^5$ data) we find a set of solutions with $\chi^2_{\rm dof}=1.69-1.81$ which allows us to assess the systematic uncertainty of the approach.
\end{abstract}

\maketitle   

{
\footnotesize
\setlength{\parskip}{-1pt}
\tableofcontents
\setlength{\parskip}{8pt}
}

\section{Introduction}
\label{sec:introduction}

Our knowledge of the baryon spectrum, as determined from analyses of data, has advanced rapidly~\cite{Zyla:2020zbs} over the past decade. The progress has been most significant for non-strange baryons, largely due to the wealth of new and more precise measurements made at electron accelerators worldwide. A substantial number of these new measurements have been performed at Jefferson Lab (JLab)~\cite{Aznauryan:2009mx} using the CLAS and Hall A detectors, at MAMI~\cite{Achenbach:2017pse},  at ELSA~\cite{Beck:2017wkb} with the Crystal Barrel detector, and also at the BESIII~\cite{Asner:2008nq} and LEPS~\cite{Shiu:2017wiw} experiments. Further investigations of the baryon spectrum are planned or realized, e.g., at J-PARC~\cite{Ohnishi:2019cif}, BGO-OD at ELSA~\cite{Alef:2019imq}, and the 12-GeV upgrade at JLab allowing studies of the electroproduction of baryon resonances to large four-momentum transfer~\cite{Carman:2020qmb, Brodsky:2020vco}.

Partial-wave analysis provides the link between large-scale experimental programs and theory approaches that focus on the intermediate-energy region, where quark confinement manifests itself in a rich spectrum of resonances. Improved and extended techniques are necessary to further our understanding of baryon structure and, in particular, to help resolve the missing-resonance problem~\cite{Koniuk:1979vw}. More generally, the baryon resonance spectrum is tightly related to the open issue of structure formation in QCD, which is arguably the least understood part of the so successful Standard Model of particle physics. Furthermore, partial-wave analysis provides the bridge to compare experiment with theories and models such as Quark Models~\cite{Isgur:1978xj,Capstick:1986bm, Capstick:1992uc, Capstick:1993kb,Ronniger:2011td, Ramalho:2011ae,Jayalath:2011uc,  Aznauryan:2012ec, Golli:2013uha, Obukhovsky:2019xrs, Ramalho:2019koj, Ramalho:2020nwk}, Dyson-Schwinger and related approaches~\cite{Roberts:1994dr,Roberts:2007ji, Eichmann:2009qa, Wilson:2011aa, Chen:2012qr, Eichmann:2012mp, Xu:2015kta, Segovia:2015hra, Eichmann:2016yit, Eichmann:2016hgl, Burkert:2017djo, Chen:2018nsg, Qin:2018dqp,Qin:2019hgk, Chen:2019fzn, Lu:2019bjs}, Roy-Steiner equations~\cite{Hoferichter:2015hva}, chiral perturbation theory (ChPT) with $\Delta$-resonance~\cite{Hemmert:1997ye, Fettes:2000bb, Long:2009wq, Navarro:2019iqj} or perturbative calculations using the complex-mass scheme~\cite{Hilt:2017iup, Bauer:2014cqa}, and chiral unitary calculations~\cite{Ruic:2011wf, Mai:2012wy, Bruns:2020lyb, Borasoy:2007ku, Meissner:1999vr, Kolomeitsev:2003kt, GarciaRecio:2003ks, Doring:2005bx, Doring:2007rz, Doring:2010fw, Doring:2009qr, Borasoy:2006sr, Gasparyan:2010xz, Garzon:2012np,Khemchandani:2013hoa, Wu:2010vk}. For example, in Refs.~\cite{Ruic:2011wf,Mai:2012wy,Bruns:2020lyb,Borasoy:2007ku,Meissner:1999vr} a gauge-invariant implementation of the full Bethe-Salpeter equation was used to fit and predict light S-wave baryons.
Specifically for pion electroproduction, ChPT has been successfully applied in the analysis of the threshold region~\cite{Bernard:1992ms, Bernard:1992rf, Bernard:1993bq, Bernard:1996bi}.
Furthermore, the spectrum of excited baryons has become accessible in lattice QCD calculations~\cite{Alexandrou:2015hxa, Alexandrou:2013ata, Engel:2013ig, Dudek:2012ag, Edwards:2012fx, Edwards:2011jj,Bulava:2010yg, Durr:2008zz, Burch:2006cc, Alexandrou:2008tn, Menadue:2011pd, Melnitchouk:2002eg}. The use of meson and baryon-type operators has also enabled the calculation of baryonic scattering amplitudes~\cite{Silvi:2021uya, Stokes:2019zdd, Andersen:2017una, Lang:2016hnn, Lang:2012db} using L\"uscher's method~\cite{Luscher:1986pf}, see also Ref.~\cite{Doring:2013glu}.
However, so far, little is known about the decay properties of baryons from such first-principles calculations.

While most of the early progress~\cite{Hohler:1984ux, Cutkosky:1979zv, Arndt:2006bf, Arndt:2009nv, Shrestha:2012ep} in baryon spectroscopy was based on the analysis of meson-nucleon scattering data, particularly pion-nucleon scattering ($\pi N\to \pi N$, $\pi N\to \pi \pi N$), photon-nucleon interactions offer the possibility of detecting unstable intermediate states with small branchings to the $\pi N$ channel~\cite{Ireland:2019uwn}. Many groups have performed either single-channel or multi-channel analyses of these photo-induced reactions. In the more recent single-channel analyses, fits have typically used isobar models~\cite{Drechsel:2007if,Tiator:2018heh, Aznauryan:2002gd,Chiang:2001as} with unitarity constraints at  lower energies, $K$-matrix-based formalisms, having built-in cuts associated with opening inelastic channels~\cite{Workman:2012jf}, and dispersion-relation constraints~\cite{Aznauryan:2002gd, Hanstein:1997tp}. Multi-channel fits have analyzed data (or, in some cases, amplitudes) from hadronic scattering data together with the photon-induced channels. These approaches have utilized unitarity more directly. The most active programs are being carried out by the Bonn-Gatchina~\cite{Anisovich:2011fc}, J\"ulich-Bonn J\"uBo~\cite{Ronchen:2015vfa}, ANL-Osaka~\cite{Kamano:2013iva}, Kent State~\cite{Shrestha:2012ep}, JPAC~\cite{Nys:2016vjz}, and Gie{\ss}en~\cite{Shklyar:2006xw} groups. At low energies, the chiral MAID analysis provides a comprehensive description of photo- and electroproduction data~\cite{Hilt:2013fda}.

With the refinement of dynamical and phenomenological coupled-channel approaches for the analysis of pseudoscalar-meson photoproduction reactions, many new states and their properties could be discovered~\cite{Zyla:2020zbs}. In this context, one should emphasize that the only model-independent definition of a resonance is by its properties in the complex energy plane. While the $Q^2$ variation of resonance couplings is expected to provide a link between perturbative QCD and the region where quark confinement sets in, so far, no unified coupled-channel analysis of photo- and electroproduction experiments exists that simultaneously studies the $\pi N$, $\eta N$ and $\Lambda K$ final states. This study provides a first step in this direction in form of an analysis of pion electroproduction data.

Going from photo- to electroproduction of pseudo-scalar meson, the number of helicity amplitudes increases from four to six, requiring more measurements for the analogous `complete experiment'~\cite{Tiator:2017cde, Dmitrasinovic:1987pc}, with a multipole decomposition adding longitudinal components to the transverse elements anchored by photoproduction analyses at $Q^2=0$. Variations of resonance couplings with $Q^2$ are expected to provide links between perturbative QCD and regions where quark confinement requires the use of lattice QCD, ChPT, or more phenomenological approaches. Exactly where this transition occurs is not precisely known. The well-known prediction~\cite{Carlson:1985mm} of an E2/M1 ratio, for the $\Delta(1232)$ state, approaching unity shows no sign of occurring, remaining essentially flat at a small negative value. See also Ref.~\cite{Pascalutsa:2006up} for a review. In contrast, other clear resonances, such as the $N(1520)$, do show rapid behavior in the low-$Q^2$ region, followed by a smooth transition to higher values of $Q^2$. The reliable determination of helicity amplitudes for $Q^2>0$ is also relevant for neutrino physics~\cite{Yao:2018pzc, Alvarez-Ruso:2015eva}. See Ref.~\cite{Nakamura:2015rta} for recent progress in this direction by the ANL/Osaka group.

Electroproduction experiments, e.g., by the CLAS Collaboration at JLab, are producing a wealth of data that, in many cases, still await a detailed analysis, in pion electroproduction~\cite{Bosted:2016spx, Bosted:2016hwk, Zheng:2016ezf, Park:2014yea} (JLab), \cite{Stajner:2017fmh} (A1 Collaboration at MAMI), $\eta$ electroproduction~\cite{Bedlinskiy:2017yxe} (JLab), \cite{Merkel:2007ig} (A1 Collaboration at MAMI), and kaon electroproduction~\cite{Achenbach:2017pse} (A2 Collaboration), \cite{Nasseripour:2008aa} (JLab). It should also be stressed that pion and kaon electroproduction experiments with the new CLAS12 detector at the 12 GeV upgrade of Jefferson Lab \cite{Brodsky:2020vco, Carman:2019lkk} will provide many data that require a timely analysis.

The ANL-Osaka group extended its dynamical coupled-channel analysis of pion electroproduction~\cite{JuliaDiaz:2009ww} to higher $Q^2$-values~\cite{Kamano:2018sfb}. Plots of the $\Delta(1232)$ amplitudes at the resonance pole position (yielding a complex amplitude) also seem to qualitatively reproduce results found for the MAID and SAID analyses~\cite{Tiator:2016btt}. However, results have generally been restricted to the low-energy $\Delta (1232)$ and $N(1440)$ states.

The most widely used single-pion electroproduction analyses, covering the resonance region, have been performed by the Mainz (MAID)~\cite{Drechsel:1998hk, Tiator:2003xr, Tiator:2003uu, Drechsel:2007if, Tiator:2011pw} and Jefferson Lab~\cite{Aznauryan:2004jd, Aznauryan:2009mx, Aznauryan:2012ba,Blin:2019fre} groups. An extensive single-pion electroproduction database, and a $K$-matrix based SAID fit, is also available~\cite{Arndt:2001si}. Eta electroproduction has been analyzed in the Eta MAID framework~\cite{Chiang:2001as}, and kaon electroproduction by the Ghent group~\cite{Corthals:2007kc}. These fits have generally utilized a Regge~\cite{Vanderhaeghen:1997ts} or Regge-plus-resonance approach~\cite{Vrancx:2014pwa} at high to medium energies. (We mention here parenthetically that the Ghent Regge approach can be improved by correctly implementing the local gauge-invariance constraints~\cite{Haberzettl:2015exa}.) Effective Lagrangian and isobar models have also been used~\cite{Mart:2002gn,Maxwell:2014txa}, with some of these available via the MAID website~\cite{MAID-web}, for both kaon and eta electroproduction~\cite{Chiang:2001as}.

These are all single-channel analyses with approaches similar to the associated real-photon fits, but generally, with the exception of the MAID and SAID analyses, not including the real-photon data as a constraint at $Q^2=0$. Both the MAID and Jefferson Lab groups have made fits using Breit-Wigner (BW) plus background models with resonance couplings extended to include a $Q^2$ dependence. We note that not all resonances can be well described by the BW form, especially if a resonance is located very close to a multiparticle threshold. In the Jefferson Lab analyses~\cite{Aznauryan:2009mx}, a further fit was again based on satisfying fixed-$t$ dispersion relation. It should be mentioned that two-pion electroproduction fits have also been performed, and compared to the single-pion results, at Jefferson Lab~\cite{Aznauryan:2005tp, Aznauryan:2011qj, Isupov:2017lnd, Mokeev:2015lda, Burkert:2019opk}. See Ref.~\cite{Aznauryan:2011qj} for a review. Remarkably, a new baryon resonance has been claimed in an analysis of combined $\pi\pi N$ photo and electroproduction~\cite{Mokeev:2020hhu}.

In this study we perform a first step towards a truly coupled-channel analysis of electroproduction data. For this, we analyse pion electroproduction data, off proton targets, in both charge channels for energies $1.13~{\rm GeV}<W<1.6~{\rm GeV}$. In this work, we do not analyze the threshold region. Mass differences within pion and nucleon multiplets, and necessary checks with ChPT~\cite{Bernard:1992ms, Bernard:1992rf, Bernard:1993bq, Bernard:1996bi} require a modified parametrization that has been developed in photoproduction~\cite{Ronchen:2014cna} but will be included for electroproduction at a later stage. The upper limit in photon virtuality for data included in the fits varies from $Q^2=4$~GeV$^2$ to $6$~GeV$^2$ to assess the stability of the solutions.

Special emphasis is put on gauge invariance and Siegert's condition~\cite{Siegert:1937yt, Tiator:2016kbr} that is manifestly included in the parametrization. The electroproduction amplitude is constructed such that at the photon point $Q^2=0$~GeV$^2$ it describes pion, $\eta$, and $K\Lambda$ photoproduction data in form of the most recent solution of the J\"ulich-Bonn analysis effort, the ``J\"uBo2017''  solution~\cite{Ronchen:2018ury}. The hadronic part of that amplitude describes also various pion-induced reactions. Extensions including $\eta$ and kaon electroproduction data, as well as the simultaneous fit of photo- and electroproduction data, are left to  future research. In this context it will be relevant to revise kaon polarization observables due to recent updates of the fundamental $\Lambda$ decay parameter $\alpha_-$~\cite{Ablikim:2018zay, Ireland:2019uja}.

This study is organized as follows. Section~\ref{sec:Preliminaries} contains formal aspects of electroproduction (kinematics, Siegert's condition, observables and multipoles), while the parametrization of electroproduction multipoles as an extension of the J\"ulich-Bonn amplitude is discussed in Sec.~\ref{sec:Formalism}. Results are presented and discussed in Sec.~\ref{sec:fits}, and the conclusions can be found in Sec.~\ref{sec:conclusions}.

\section{Preliminaries}
\label{sec:Preliminaries}

\begin{figure*}[t]
\centering
\includegraphics[width=0.6\linewidth,trim=0 7cm 12.6cm 7cm,clip]{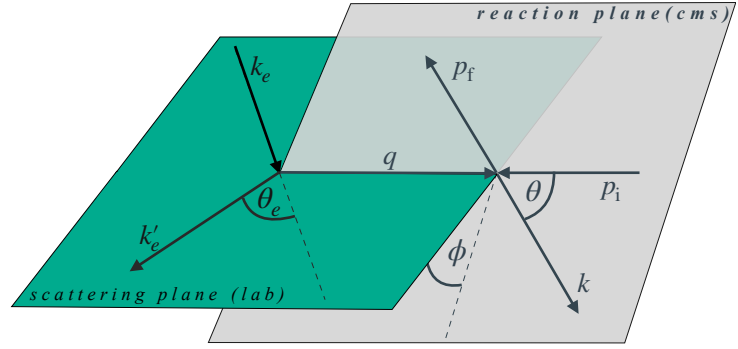}
\caption{
\label{fig:kinematics}
Kinematics of an electroproduction experiment. The scattering plane is defined by the respective in/outgoing electron momenta $k_e/k_e'$ with the electron-scattering angle $\theta_e$. The reaction plane is spanned by the virtual photon and the outgoing meson, scattered by an angle $\theta$.}
\end{figure*}

\subsection{Kinematics }
\label{subsec:kinematics}

The pion electroproduction process in question occurs via the following reaction (bold symbols denote three-momenta throughout the manuscript)
\begin{align*}
\gamma^*(\bm{q})+p(\bm{p}_{\rm i})
\to
\pi(\bm{k})+N(\bm{p}_{\rm f})\,,
\end{align*}
with the virtual photon $\gamma^*(\bm{q})$ being produced via $e_{\rm in}(\bm{k}_e)\to e_{\rm out}(\bm{k}'_e)+\gamma^*(\bm{q})$. Thus, the momentum transfer $Q^2=-\omega^2+\bm{q}^2$ is non-negative for spacelike processes, and acts as an independent kinematical variable in addition to the total energy in the center-of-mass (cms) frame, $W$. In this frame, the magnitude of the three momentum of the photon ($q=|\bm{q}|$) and produced pion ($k=|\bm{k}|$) read
\begin{align}
q=\frac{\sqrt{\lambda(W^2,m^2,-Q^2)}}{2W}\,,~
k=\frac{\sqrt{\lambda(W^2,m^2,M^2)}}{2W}\,,
\label{eq:qk}
\end{align}
where $\lambda(x,y,z)=x^2+y^2+z^2-2xy-2yz-2zx$ denotes the usual triangle K\"all\'en function. The pion and nucleon masses are denoted throughout this manuscript by $M$ and $m$, respectively. From this, the photon energy is $\omega=(W^2-m^2-Q^2)/(2W)$. The angular structure of the above process is depicted in Fig.~\ref{fig:kinematics}, where $\theta_{e}$ is the angle of the in/outgoing electron in the scattering plane, $\phi$ is the angle of the reaction plane to the scattering plane and $\theta$ is the cms meson scattering-angle in the latter plane. The world data on electroproduction is represented with respect to the five variables
\begin{align*}
\mathcal{O}(Q^2,W,\phi,\theta,\epsilon)\,,
\end{align*}
where $\epsilon=[1+2q_L^2/Q^2\tan^2(\theta_e/2)]^{-1}$ and $\mathcal{O}$
denotes observables discussed in the next section. Here, $q_L$ is the photon three-momentum in the lab frame.

There are several important limits/thresholds of this kinematics which will be discussed later and thus require an introduction.
\begin{itemize}
    \item \emph{Photon point}: This corresponds to $Q^2\equiv0$, reducing the process to pion photoproduction. In this limiting case, denoted by a subscript $\gamma$, the amplitudes are independent of the angle $\phi$ and
        \begin{align}
        \omega_\gamma=q_\gamma=\frac{W^2-m^2}{2W}\,.
        \label{eq:qg}
        \end{align}
        The electroproduction amplitudes are constrained by photoproduction data via the multipoles of the JüBo model, see Sec.~\ref{subsec:jubo}.
    \item \emph{Production threshold}: This refers to the lowest physical energy of the final meson-baryon pair, i.e., $W=(m+M)~\Leftrightarrow k\equiv0$.
    \item  \emph{Pseudo-threshold}: This denotes an unphysical point $q=0$ or correspondingly
  \begin{align}
        Q^2_{\rm PT\pm}=-(W\pm m)^2\,.
        \end{align}
    The particular importance of this kinematic point arises from Siegert's condition, which will serve as a boundary condition for our parametrization of the multipoles, see Sec.~\ref{sec:Formalism}.
\end{itemize}

\subsection{Transition amplitudes -- multipoles and Siegert's condition }
\label{subsec:Transition-amplitudes}

For a general photon-induced photo- or electroproduction  of a meson off a nucleon, the transition amplitude takes the form
\begin{align}
    T_{\rm fi}=
    &\langle {\rm f}|(-i\int d^4r A^{\mu}J_{\mu})|{\rm i}\rangle\,,
\end{align}
with $J_\mu$ denoting the electromagnetic current, see, e.g., the seminal papers~\cite{Dennery:1961zz,Berends:1967vi,CiofiDegliAtti:1981wk} for more details. The components of the vector potential $A^\mu$ are solutions of the Laplace equation. For instance, for the time component (scalar potential),
\begin{align}
    \left(\nabla^2+q^2\right)A^0=0\,.
\end{align}
Then, $A^0$ can be decomposed into contributions with given angular momentum values,
\begin{align}
    A^0=\sum_{j_{\gamma}m_\gamma}\int\frac{dq}{2\pi}\frac{q^2}{ \sqrt{\omega}}
    (a_{j_{\gamma}m_\gamma}u^C_{j_{\gamma}m_\gamma}
    +a^\dagger_{j_{\gamma}m_\gamma}u^{C*}_{j_{\gamma}m_\gamma})\,,
\end{align}
where $j_\gamma$ is the angular momentum of the photon and $a_{j_{\gamma}m_\gamma}^{(\dagger)}\equiv a_{j_{\gamma}m_\gamma}^{(\dagger)}(q)$ are annihilation  (creation) operators. Furthermore,
\begin{align}
    u^C_{j_{\gamma}m_\gamma}\equiv u^C_{j_{\gamma}m_\gamma}({\bm q},{\bm r})=j_{j_\gamma}(qr)Y_{j_{\gamma}m_\gamma}(\theta,\phi)\,.
\label{eq:uC}
\end{align}
Everywhere, $j_{j_\gamma}(qr)$ and $Y_{lm}(\theta,\phi)$ denote the spherical Bessel functions and spherical harmonics, respectively.

Similarly, the three-vector potential can be expanded in elementary vectors ${\bm u}^x$ with a given angular momentum as in the scalar case,
\begin{align}
    {\bm A}=
    \sum_{j_{\gamma}m_\gamma}\sum_{x}
    \int{dq\over2\pi}{q^2\over \sqrt{\omega}}
    &\Big(
    a_{j_{\gamma}m_\gamma}^x(q){\bm u}^x_{j_{\gamma}m_\gamma}({\bm q},{\bm r})\\
    &\phantom{-}+a^{x\dagger}_{j_{\gamma}m_\gamma}(q){\bm u}^{x*}_{j_{\gamma}m_\gamma}({\bm q},{\bm r})
    \Big)\,,\nonumber
\end{align}
where $\omega$ is the photon energy, while $x\in\{E,M,L\}$ labels electric, magnetic and longitudinal  components, respectively. The vectors ${\bm u}^x$ can be constructed from the scalar $u^C$ using differential operators,
\begin{align}
    {\bm u}^L_{j_{\gamma}m_\gamma}({\bm q},{\bm r})&=\frac{i}{q}\nabla u^C_{j_\gamma m_\gamma}({\bm q},{\bm r}) \ ,\nonumber\\
    {\bm u}^M_{j_{\gamma}m_\gamma}({\bm q},{\bm r})&=\frac{{\bm r}\times \nabla}{  \sqrt{j_\gamma(j_\gamma-1)}}u^C_{j_{\gamma}m_\gamma}({\bm q},{\bm r}) \ ,\nonumber\\
    {\bm u}^E_{j_{\gamma}m_\gamma}({\bm q},{\bm r}) &= \frac{\nabla\times \bm L}{q \sqrt{j_\gamma(j_\gamma+1)}}u^C_{j_{\gamma}m_\gamma}({\bm q},{\bm r})\,.
\label{eq:uELM}
\end{align}
Because $\nabla$, ${\bm L}=-i{\bm r}\times\nabla$, and $\nabla\times{\bm L}$ commute with the Laplacian, ${\bm u}^L$, ${\bm u}^M$ and ${\bm u}^E$  are also solutions of the Laplace equation, and are orthogonal to each other. The corresponding Coulomb ($C$), longitudinal ($L$), magnetic ($M$) and electric ($E$) multipoles then read for a given photon angular momentum state
\begin{align}
    C_{j_{\gamma}m_\gamma}({\bm q})&=\langle {\rm f}\vert\int d^3r\, u^C_{j_{\gamma}m_\gamma}({\bm q},{\bm r})\rho({\bm r})\vert {\rm i}\rangle\,,\nonumber\\
    L_{j_{\gamma}m_\gamma}({\bm q})&=\langle {\rm f}\vert\int d^3r\, {\bm u}^L_{j_{\gamma}m_\gamma}({\bm q},{\bm r})\cdot{\bm J}({\bm r})\vert {\rm i}\rangle\,,\nonumber\\
    E_{j_{\gamma}m_\gamma}({\bm q})&=\langle {\rm f}\vert\int d^3r\, {\bm u}^E_{j_{\gamma}m_\gamma}({\bm q},{\bm r})\cdot{\bm J}({\bm r})\vert {\rm i}\rangle\,,\nonumber\\
    M_{j_{\gamma}m_\gamma}({\bm q})&=\langle {\rm f}\vert\int d^3r\, {\bm u}^M_{j_{\gamma}m_\gamma}({\bm q},{\bm r})\cdot{\bm J}({\bm r})\vert {\rm i}\rangle\,,
\label{eq:general-multipoles}
\end{align}
respectively. Only three multipoles here are independent because of the continuity equation for the current, thus relating scalar and longitudinal multipoles according to
\begin{align}
\omega C_{j_\gamma}({\bm q})=q L_{j_\gamma}({\bm q})\,,
\label{eq:C-to-L-relation}
\end{align}
where the $m_\gamma$ dependence of the multipoles cancels out by the Wigner-Eckart theorem, see, e.g., Ref.~\cite{CiofiDegliAtti:1981wk}.

Finally, and following Ref.~\cite{CiofiDegliAtti:1981wk}, we note that in the long-wavelength limit ($q\to0$) (pseudo-threshold) $j_{j_\gamma}(qr)\to (qr)^{j_{\gamma}}/(2{j_\gamma}+1)!!$. Using then Eq.~\eqref{eq:uC} and $\nabla\times\bm{L}(r^{j_\gamma} Y_{j_{\gamma}m_\gamma})=i({j_\gamma}+1)\nabla(r^{j_\gamma}Y_{j_{\gamma}m_\gamma})$ yields straightforwardly
\begin{align}
    {\bm u}^E_{j_\gamma}({\bm k},{\bm r})
    \to \frac{i}{q}\sqrt{\frac{{j_\gamma}+1}{j_\gamma}}\nabla u^C_{j_\gamma}({\bm q},{\bm r})\,.
\end{align}
Then, substituting this into Eqs.~\eqref{eq:uELM} and \eqref{eq:general-multipoles} results in an exact relation between electric and longitudinal multipole at the pseudo-threshold,
\begin{equation}
    E_{j_\gamma}
 =
\sqrt{\frac{{j_\gamma}+1}{j_\gamma}} L_{j_\gamma}~,\quad\text{at}~q=0~,
\label{eq:SC-j-basis}
\end{equation}
called Siegert's theorem that provides an important constraint on our parametrization of the electroproduction multipoles. The practical implementation of it will be discussed in the next section.

\subsection{Transition amplitudes -- CGLN and helicity amplitudes }
\label{subsec:multipole}

To find a practical access to the ($ELM$) multipoles~\eqref{eq:general-multipoles}, introduced above, we follow the seminal works~\cite{Dennery:1961zz,Chew:1957tf} utilizing the nomenclature of Ref.~\cite{Berends:1967vi}. Taking the $z$-axis as the quantization axis and working in the center of mass of the final pion-nucleon system which yields the general Lorentz covariant transition matrix element
\begin{align}
T_{\rm fi}=8\pi W\,\chi^\dagger_{\rm f}
\sum_{a=1}^{6}\Big(\mathcal{F}_a \mathcal{G}_a \Big)
\chi_{\rm i}\,.
\label{eq:transitions-CGLN}
\end{align}
Note that charge conservation is already implemented here thus reducing the number of independent structures to six. Here, $\chi$ denotes the two-component spinor, and $\mathcal{F}_a$ are the CGLN amplitudes\footnote{Referring to the authors of Ref.~\cite{Chew:1957tf} (Chew, Goldberger, Low, Nambu) those amplitudes were originaly derived for the photoproduction amplitudes, but extended later by Dennery~\cite{Dennery:1961zz} to electroproduction.},
being coefficients of
\begin{align}
\mathcal{G}=\{
  & i({\bm\sigma}\cdot{\bm a}),\,
    ({\bm\sigma}\cdot\hat{\bm k})\,
    ({\bm\sigma}\cdot(\hat{\bm q}\times{\bm a})),\,\nonumber\\
&   i({\bm\sigma}\cdot\hat{\bm q})(\hat{\bm k}\cdot{\bm a}),\,
    i({\bm\sigma}\cdot\hat{\bm k})(\hat{\bm k}\cdot{\bm a}),\,\nonumber\\
&   i({\bm\sigma}\cdot\hat{\bm q})(\hat{\bm q}\cdot{\bm a}),\,
    i({\bm\sigma}\cdot\hat{\bm k})(\hat{\bm q}\cdot{\bm a})
\}\,
\end{align}
with $a_\mu=\epsilon_\mu-(\epsilon_0/\omega)q_\mu$ ($\epsilon_0$ being the 0-component of the polarization vector $\epsilon_\mu$~\cite{Dennery:1961zz}), and hat denoting the normalization of a respective three-vector.

The six types of transitions~\eqref{eq:transitions-CGLN} can be related to the eigenamplitudes of definite parity and relative angular momentum of the pion-nucleon pair $(\ell)$~\cite{Pearlstein:1957zz}. These amplitudes are identified with electric $E_{\ell\pm}$, magnetic $M_{\ell\pm}$ and scalar or Coulomb multipoles~\eqref{eq:C-to-L-relation}, see Eqs.~\eqref{eq:general-multipoles}.  Expanding with respect to the Legendre polynomials $P_\ell(\cos{\theta})$ yields
\begin{align}
    \mathcal{F}_1 &=\sum_{\ell\ge 0}\left(\left( \ell M_{\ell+} + E_{\ell+}
    \right) P'_{\ell+ 1}\right.\nonumber\\
    &\left.~~~~~~~~~~~~~~~~~~~~~~~~
    +\left( (\ell+1) M_{\ell-}+E_{\ell-}
    \right) P'_{\ell- 1}\right)\,,\nonumber\\
    \mathcal{F}_2 &=\sum_{\ell\ge 1}\left((\ell+ 1) M_{\ell+} + \ell M_{\ell
    -}\right)P'_{\ell}\,,
    \nonumber\\
    \mathcal{F}_3 &=\sum_{\ell\ge 1}\left(\left( E_{\ell+} - M_{\ell+}
    \right) P''_{\ell+ 1} +\left( E_{\ell-} + M_{\ell-} \right)
    P''_{\ell- 1}\right)\,,
    \nonumber\\
    \mathcal{F}_4 &=\sum_{\ell\ge 2}\left(M_{\ell+} - E_{\ell+} -  M_{\ell-}
    - E_{\ell-}\right)P''_{\ell}\,,
    \nonumber\\
    \mathcal{F}_5 &=\sum_{\ell\ge 0} \left((\ell+ 1) L_{\ell+} P'_{\ell+1}
    -\ell\; L_{\ell-} P'_{\ell- 1}\right)\,,
    \nonumber\\
    \mathcal{F}_6 &=\sum_{\ell\ge 1} \left(\ell L_{\ell- } - (\ell+ 1)
    L_{\ell+}\right)P'_{\ell}\,.
\label{eq:F_amplitudes}
\end{align}
Here we have suppressed the dependence of the multipoles on $W$ and $Q^2$ for simplicity. The total angular momentum is given by $J=\ell\pm1/2=\ell\pm$. The multipole decomposition yields a certain behavior of the multipoles at the physical and pseudo-threshold,
\begin{align}
\begin{tabular}{llllll}
           &   &            &           & $k\to 0$~~ & $q\to 0$\\
\hline
$E_{\ell+}$&and&$L_{\ell+}$ &$\text{for~}\ell\geq0$~~~~& $k^\ell$  & $q^\ell$\rule{0pt}{2.6ex}\\
$L_{\ell-}$&   &            &$\text{for~}\ell=1$   ~~& $k$       & $q$\\
$M_{\ell+}$&and&$M_{\ell-}$ &$\text{for~}\ell\geq1$~~& $k^\ell$  & $q^\ell$\\
$E_{\ell-}$&and&$L_{\ell-}$ &$\text{for~}\ell\geq2$~~& $k^\ell$  & $q^{\ell-2}$\\
\hline
\end{tabular}
\label{eq:pseudo-thre-conditions}
\end{align}

Finally, coming back to Siegert's theorem~\eqref{eq:SC-j-basis}, we note that the multipoles in the previous section are labeled with the incident photon angular momentum, $j_\gamma$, while in the current section they are indexed by the final-state orbital angular momentum $\ell$. Transforming to the latter basis Siegert's theorem takes the form~\cite{Tiator:2016kbr}
\begin{equation}
    \frac{E_{\ell+}}{L_{\ell+}}=1
\text{~~and~~}
    \frac{E_{\ell-}}{L_{\ell-}}=\frac{\ell}{1-\ell}~,
    \text{~~at~~$q=0$}\ ,
\label{eq:Siegerts_condition}
\end{equation}
which is also the form employed in this work at $Q^2=Q^2_{{\rm PT}-}$, referring to it as  Siegert's condition\footnote{We do not employ the same condition at the second pseudo-threshold, since it is located much further away from the physical region.}. This is crucial, since the current parametrization relies on the continuation of the available photoproduction solution to $Q^2>0$, see Sec.~\ref{sec:Formalism}. The latter, however, do not restrict the longitudinal multipoles. Equating the latter to the electric multipole at the pseudo-threshold provides a solution to this problem and is employed in this work.

\subsection{Response functions and observables }
\label{subsec:observables}

Free parameters for the various multipoles will be determined by fits to data of  differential cross sections and other observables. These observables are written in terms of response functions, $R(W,Q^2,\theta)$ which can be related to the transition amplitudes, conveniently employing the helicity formalism~\cite{Jacob:1959at}. In particular, the differential cross section is a function of five kinematic variables discussed in Sec.~\ref{subsec:kinematics} ($W,Q^2,\theta,\phi,\epsilon$) defined as
\begin{align}
    \frac{d\sigma}{d\Omega_{\rm f}dE_{\rm f}d\Omega}&=
    \left(
    \frac{\alpha}{2\pi^2}
    \frac{E_{\rm f}}{E_{\rm i}}
    \frac{q_L}{Q^2}
    \frac{1}{1-\epsilon}\right)
    \frac{d\sigma^v}{d\Omega}\,,
\label{eq:diff-cross-sec.}
\end{align}
where $\Omega$ refers to the angles of the final meson baryon system ($\theta$, $\phi$) and $\Omega_{\rm f}$ are the angles of the final electron at energy $E_{\rm f}$. The differential cross section $d\sigma^v/d\Omega$ for the virtual photon sub-process is commonly further decomposed as
\begin{align}
\label{eq:DSG}
    \frac{d\sigma^v}{d\Omega}=&
    \sigma_{T}+\epsilon\sigma_{L}
    +\sqrt{2\epsilon(1+\epsilon)}\sigma_{LT}\cos{\phi}\\
    &+\epsilon\sigma_{TT}\cos{2\phi}
    +h\sqrt{2\epsilon(1-\epsilon)}\sigma_{LT'}\sin{\phi}
\,,\nonumber
\end{align}
where $h$ is the helicity of the incoming electron.
The quantities $\{\sigma_x|x=(T,TT,LT,L,LT')\}$ are referred to as structure functions. Data involving polarized quantities are included from Jefferson Lab experiments: (1) From Ref.~\cite{Joo:2003uc,Joo:2004mi} via longitudinal-transverse structure functions~$\sigma_{LT'}$ and (2) The  $K_{1D}=\{K_{1D}^X|X=A,B,..,T\}$ observables from Ref.~\cite{Kelly:2005jj} related to the response functions as shown in  Tab.~\ref{tab:KE05-conversion} in Appendix~\ref{app:sec:Kelly}. In general, and following the convention of Ref.~\cite{Tiator:2017cde}, the structure functions in Eq.~\eqref{eq:DSG} are obtained from the response functions $R$. The latter denote the coefficients that expand the azimuthal angle dependence of the general differential cross section of an electroproduction reaction when all polarizations are taking into account, see e.g., Ref.~\cite{Tiator:2017cde}. In our case, and using  $q_\gamma\equiv q(Q^2=0)$ from Eq.~\eqref{eq:qg}, the required connection reduces to
\begin{align}
    &\sigma_T=\frac{k}{q_\gamma}R^{00}_T\,,~~
    \sigma_L=\frac{k}{q_\gamma}\frac{Q^2}{\omega^2}R^{00}_L\nonumber\,,~~
    \sigma_{TT}=\frac{k}{q_\gamma}R^{00}_{TT}\\
    &\sigma_{LT}=\frac{k}{q_\gamma}\frac{\sqrt{Q^2}}{\omega}R^{00}_{LT}\,,~~
    \sigma_{LT'}=\frac{k}{q_\gamma}\frac{\sqrt{Q^2}}{\omega}R^{00}_{LT'}\,.
\label{eq:structure-functions}
\end{align}
The response functions can be expressed in terms of helicity amplitudes ($H$),
\begin{align}
    &R^{00}_T~\,=\left(|H_1|^2+|H_2|^2+|H_3|^2+|H_4|^2\right)/2\,,\nonumber\\
    &R^{00}_L~\,=\left(|H_5|^2+|H_6|^2\right)\,,\nonumber\\
    &R^{00}_{LT}\,=\left((H_1-H_4)H_5^*+(H_2+H_3)H_6^*\right)/\sqrt{2}\,,\nonumber\\
    &R^{00}_{TT}\,=\Re\left(H_3H_2^*-H_4H_1^*\right)\,,\nonumber\\
    &R^{00}_{LT'}=\Im\left((H_4-H_1)H_5^*-(H_2+H_3)H_6^*\right)/\sqrt{2}\,,\nonumber\\
    &R_{LT}^{0Y} = -\Re\left( (H_2+H_3)H^*_5+(H_4-H_1)H^*_6 \right)/\sqrt{2}\,,
    \label{eq:response-functions}
\end{align}
where where the connection to CGLN $\mathcal{F}$ amplitudes is given by
\begin{align}
    H_1 &= \sin\theta\cos\nicefrac{\theta}{2} (-\mathcal{F}_3 -\mathcal{F}_4 )/\sqrt{2}\,,
    \nonumber\\
    H_2 &= \sqrt{2} \cos\nicefrac{\theta}{2} \left( \mathcal{F}_2 - \mathcal{F}_1 + ( \mathcal{F}_3 - \mathcal{F}_4) \sin^2 \nicefrac{\theta}{2} \right)\,,
    \nonumber\\
    H_3 &=\sin \theta \sin \nicefrac{\theta}{2} (\mathcal{F}_3 -\mathcal{F}_4)/\sqrt{2}\,,
    \nonumber\\
    H_4 &= \sqrt{2} \sin \nicefrac{\theta}{2} \left(\mathcal{F}_1 + \mathcal{F}_2 + (\mathcal{F}_3 + \mathcal{F}_4)\cos^2 \nicefrac{\theta}{2} \right)\,,
    \nonumber\\
    H_5 &= \cos \nicefrac{\theta}{2} (\mathcal{F}_5 + \mathcal{F}_6 ) \,,
    \nonumber\\
    H_6 &= \sin \nicefrac{\theta}{2} (\mathcal{F}_6 - \mathcal{F}_5) \,,
    \label{eq:Helicity-amplitudes}
\end{align}
following again the phase convention of Refs.~\cite{Jacob:1959at,Walker:1968xu}.
Several other observables are included in the fits as described in Sec.~\ref{subsec:exp-data},
\begin{align}
\label{eq:polarization_obs}
P_Y      &= -\sqrt{2\epsilon(1+\epsilon)}\frac{\omega}{\sqrt{Q^2}} \frac{R_{LT}^{0Y}}{R_{T}^{00}+\epsilon\omega^2/Q^2 R_{L}^{00}}\,,\nonumber\\
\rho_{LT}&= \phantom{-}\sqrt{2\epsilon(1+\epsilon)}\frac{R_{LT}^{00}}{R_{T}^{00}+\epsilon(R_{L}^{00}+R_{TT}^{00})}\,,
\nonumber\\
\rho_{LT'}  &= \phantom{-}\sqrt{2\epsilon(1-\epsilon)}\sin\phi   \frac{\sigma_{LT'}}{d\sigma^v/d\Omega}\,,
\end{align}
while the $K_{D1}$-observables are given explicitly in Appendix~\ref{app:sec:Kelly}.

\section{Multipole parametrization}
\label{sec:Formalism}

\subsection{The J\"ulich-Bonn dynamical coupled-channel approach }
\label{subsec:jubo}

The input at the photon point is provided by the J\"ulich-Bonn (J\"uBo) framework, a dynamical coupled-channel approach that aims at the extraction of the nucleon resonance spectrum in a combined analysis of pion- and photon-induced hadronic reactions. In this approach, two-body unitarity and analyticity are respected and the baryon resonance spectrum is determined in terms of poles in the complex energy plane on the second Riemann sheet. A detailed description of the model can be found in Refs.~\cite{Ronchen:2012eg,Ronchen:2014cna} and references therein.

The purely hadronic scattering process of a meson-baryon pair $\nu$ is described in a field-theoretical framework by potentials $V_{\mu\nu}$, that are derived from a chiral Lagrangian and iterated in a Lippmann-Schwinger equation
\begin{align}
T_{\mu\nu}&(k,p',W)=V_{\mu\nu}(k,p',W)
\label{eq:juboscat}\\
&\hspace{-3ex}\mbox{}+\sum_\kappa\int\limits_0^\infty dp\,p^2\,V_{\mu\kappa}(k,p,W)G^{}_\kappa(p,W)\,T_{\kappa\nu}(p,p',W)\, ,
\nonumber
\end{align}
where the indices $\mu$, $\nu$ and $\kappa$ denote the outgoing, incoming and intermediate meson-baryon channels, respectively. The model incorporates the two-body channels $\pi N$, $\eta N$, $K\Lambda$, and  $K\Sigma$ and the channels $\rho N$, $\sigma N$ and $\pi\Delta$, which effectively parameterize the $\pi\pi N$ channel. In Eq.~(\ref{eq:juboscat}), $k$ ($p'$) indicates the modulus of the outgoing (incoming) three-momentum in the cm system, which can be on- or off-shell. The propagator $G_\kappa$ is given by
\begin{align}
G_\kappa(p,W)=\frac{1}{W-E_a(p)-E_b(p)+i\epsilon}\,,
\label{eq:juboprop}
\end{align}
with the on-mass-shell energies $E_a=\sqrt{m_a^2+p^2}$ and  $E_b=\sqrt{m_b^2+p^2}$  of the intermediate particles $a$ and $b$ in channel $\kappa$ with masses $m_a$ and $m_b$. While Eq.~(\ref{eq:juboprop}) applies to the channels $\kappa=\pi N$, $\eta N$, $K\Lambda$, or $K\Sigma$, the propagator is of a more complex form for channels with unstable particles, i.e. $\rho N$, $\sigma N$ and $\pi\Delta$~\cite{Doring:2009yv,Krehl:1999km}. The scattering potential $V_{\mu\nu}$ is constructed from $s$-channel processes that account for genuine resonances, $t$- and $u$-channel exchanges of mesons and baryons, and contact diagrams that are included to absorb physics beyond the explicitly included processes.

The photoproduction process is described in the semi-phenomenological approach of Ref.~\cite{Ronchen:2014cna}, where the electric or magnetic photoproduction multipole amplitude (see Eq.~\ref{eq:general-multipoles}) is given by
\begin{align}
{\cal M}_{\mu\gamma}(k,W)&=V_{\mu\gamma}(k,W)
\label{eq:jubophoto}
 \\
&\hspace{-7ex}\mbox{}+\sum_\kappa\int\limits_0^\infty dp\,p^2\,T_{\mu\kappa}(k,p,W)G^{}_\kappa(p,W)V_{\kappa\gamma}(p,W)\,.
\nonumber
\end{align}
Here, the index $\gamma$ denotes the initial $\gamma N$ channel and $\mu$ ($\kappa$) the final (intermediate) meson-baryon pair, while $T_{\mu\kappa}$ is the hadronic half-off-shell matrix of Eq.~(\ref{eq:juboscat}) and $k$ denotes, again, the momentum of the outgoing meson.

The photoproduction kernel $V_{\mu\gamma}$ is constructed as
\begin{align}
V_{\mu\gamma}(p,W)=\alpha^{\rm NP}_{\mu\gamma}(p,W)+\sum_{i} \frac{\gamma^a_{\mu;i}(p)\,\gamma^c_{\gamma;i}(W)}{W-m_i^b} \ ,
\label{eq:jubovg}
\end{align}
where $\gamma^c_{\gamma;i}$ describes the interaction of the photon with the resonance state $i$ with bare mass $m_i^b$ and $\alpha^{\rm NP}_{\mu\gamma}$ accounts for the coupling of the photon to the so-called background or non-pole part of the amplitude. Both quantities are parametrized by energy-dependent polynomials. See Appendix~\ref{app:sec:photo_details} for details. In particular, we note that the hadronic resonance vertex function $\gamma^a_{\mu;i}$ in Eq.~(\ref{eq:jubovg}) is the same as in the hadronic scattering potential to ensure the cancellation of the poles in Eq.~(\ref{eq:jubovg}). Explicit expressions for $\gamma^a_{\mu;i}$ can be found in Ref.~\cite{Doring:2010ap}.

The hadronic scattering potential $V_{\mu\nu}$ and the polynomials in $\gamma^c_{\gamma;i}$ and $\alpha^{\rm NP}_{\mu\gamma}$ contain free parameters that are fitted to the data in a $\chi^2$ minimization using MINUIT on the JURECA supercomputer at the J\"ulich Supercomputing Centre~\cite{jureca}. In its most recent form~\cite{Ronchen:2018ury} the J\"uBo model describes the reactions $\pi N\to \pi N$, $\eta N$, $K\Lambda$ and $K\Sigma$ in addition to pion, eta and $K^+\Lambda$ photoproduction off the proton. More than 48,000 data points were analyzed in simultaneous fits and the $N$ and $\Delta$ spectrum was determined.

\subsection{Extension of the J\"uBo formalism to electroproduction }
\label{subsec:JuBo extended}

To include electroproduction reactions in the J\"uBo formalism, the photoproduction formalism outlined in the previous section is extended to handle virtual photons with $Q^2>0$. Following Eq.~(\ref{eq:jubophoto})  we first introduce a generic function ($\bar{\cal{M}}$) for each electromagnetic multipole  (${\cal M}_{\mu\gamma^*}\in \{E_\mu,L_\mu,M_\mu\}$) as
\begin{align}
   \bar{\cal M}_{\mu\gamma^*}&(k,W,Q^2)=V_{\mu\gamma^*}(k,W,Q^2)
       \label{eq:m_electro}\\
   &\hspace{-6ex}\mbox{}+\sum_\kappa\int\limits_0^\infty dp\, p^2\, T_{\mu\kappa}(k,p,W)G_\kappa(p,W)V_{\kappa\gamma^*}(p,W,Q^2)\,,
\nonumber
\end{align}
with $\kappa\in\{\pi N,\eta N, K\Lambda, K\Sigma, \pi\Delta, \rho N\}$ and $\gamma^*$ denoting the ingoing $\gamma^*N$ state. The electroproduction kernel $V_{\mu\gamma^*}$ in Eq.~\eqref{eq:m_electro} is parametrized as
\begin{align}
    V_{\mu\gamma^*}(p,W,Q^2)&=
    \alpha^{NP}_{\mu\gamma^*}(p,W,Q^2)
    \nonumber\\
    &\hspace{-5ex}\mbox{}+\sum_{i=1}^{i_\text{max}}{\gamma^a_{\mu;i}(p)\gamma^c_{\gamma^*;i}(W,Q^2)\over W-m^b_i}\,,
    \label{eq:v_electro}
\end{align}
introducing the $Q^2$-dependence via a separable ansatz,
\begin{align}
    \alpha^{NP}_{\mu\gamma^*}(p,W,Q^2)& ={\tilde F}^\mu(Q^2)\alpha^{NP}_{\mu\gamma}(p,W)\nonumber\\
    \gamma^c_{\gamma^*;i}(W,Q^2)&       ={\tilde F}_i(Q^2)\gamma^c_{\gamma;i}(W)\,,
    \label{eq:ff_electro_2}
\end{align}
with a channel-dependent form-factor ${\tilde F}^\mu(Q^2)$ and another channel-independent form-factor ${\tilde F}_i(Q^2)$ that depends on the resonance number $i$. Note that the channel-dependence is inherited from the structure of the photoproduction ansatz of Eq.~(\ref{eq:jubovg}), which separates the photon-induced vertex ($\gamma^c$) from the decay vertex of a resonance to the final meson-baryon pair ($\gamma_\mu^a$).

Both ${\tilde F}^\mu(Q^2)$ and ${\tilde F}_i(Q^2)$  are chosen as
\begin{align}
    \label{eq:formfactor-Ftilde}
    {\tilde F}(Q^2)={\tilde F}_D(Q^2)\,e^{-\beta_0 Q^2/ m^2}\,P^N(Q^2/m^2)\,,
\end{align}
where
\begin{align}
    \label{eq:formfactor-FD}
    {\tilde F}_D(Q^2)=\frac{1}{(1+Q^2/b^2)^{2}}\,
    \frac{1+e^{-Q_r^2/Q_w^2}}{1+e^{(Q^2-Q_r^2)/Q_w^2}}
\end{align}
is a combination of the empirical dipole form-factor with $b^2=0.71$~GeV${}^2$, usually implemented in such problems, see, e.g., Ref.~\cite{Scadron:2007qd}, as well as a Woods-Saxon form factor with $Q_w^2=0.5~{\rm GeV}^2$ and  $Q_r^2=4.0~{\rm GeV}^2$, which is introduced to ensure that at large $Q^2$ the multipoles vanish sufficiently rapidly. Furthermore, the polynomial $P^N(x)= 1 + \beta_1x +\beta_2x^2 +...+ \beta_Nx^N$ is to be fitted to data, along with the parameter $\beta_0$. A similar parametrization is chosen in Refs.~\cite{Kamano:2018sfb, Sato:2000jf}.

For electric and magnetic multipoles, the quantities $\gamma^a_{\mu;i}$, $\alpha^{NP}_{\mu\gamma}$, $\gamma^c_{\gamma;i}$, and $m_i^b$ in the electroproduction amplitude of Eqs.~\eqref{eq:v_electro} and \eqref{eq:ff_electro_2}, as well as $T_{\mu\kappa}$ in Eq.~\eqref{eq:m_electro}, represent the input at the photo-point in the current analysis. Numerical values are taken from the J\"uBo2017 solution of Ref.~\cite{Ronchen:2018ury}.

For longitudinal multipoles there is no information on $\alpha_{\mu\gamma^*}^{NP}$ or $\gamma_{\gamma^*;i}^c$ at the photo-point. To overcome this we employ the following strategy.

{\bf 1)} Following Siegert's condition~\eqref{eq:Siegerts_condition} we apply
\begin{align}
\alpha^{NP}_{L_{\ell\pm}}(Q^2)
= \frac{\omega(Q^2)}{\omega(Q^2_{\rm PT-})}
\frac{\alpha^{NP}_{E_{\ell\pm}}(Q^2_{\rm PT-})}{{\tilde F}_D(Q^2_{\rm PT-})}
{\tilde F}_D(Q^2)D_{\ell\pm}(Q^2)
\,,
\label{pt_cond4}
\end{align}
and similarly for $\gamma_{\gamma^*;i}$. The photon energy $\omega$ was defined below Eq.~\eqref{eq:qk}. The new functions $D_{\ell\pm}(Q^2)$ incorporate Siegert's condition exactly, ensuring at the same time a $Q^2$ falloff behavior. Explicitly they read
\begin{align}
D_{\ell+}(Q^2)&=
    e^{-\beta_0 k/k_\gamma}\,    P^{N}(k/k_\gamma)\,,\nonumber\\
D_{\ell-}(Q^2)&=
    -\frac{\ell-1}{\ell}e^{-\beta_0 k/k_\gamma}\,
    P^{N}(k/k_\gamma)\,,
\label{eq:D-formfactor}
\end{align}
where $k_\gamma=k(0)$ and $k$ from Eq.~\eqref{eq:qk}.

{\bf 2)} For the two cases with vanishing electric multipole, i.e., $(\ell\pm,I)=(1-,1/2)$ and $(\ell\pm,I)=(1-,3/2)$, the longitudinal multipole is obtained from the magnetic one via
\begin{align}
\alpha^{NP}_{L_{\ell\pm}}(Q^2)&=\zeta^{NP}\frac{\omega(Q^2)}{\omega(Q^2_{\rm PT-})}{\tilde F}^\mu(Q^2)\alpha^{NP}_{M_{\ell\pm}}(p,W)\,,
\label{eq:zetanp}
\end{align}
and similarly for $\gamma_{\gamma^*;i}$. The new real-valued normalization constants $\zeta^{NP}$ will be determined from the fit.


Imposing the pseudothreshold-constraints of Table~\eqref{eq:pseudo-thre-conditions}, a $q$ and $\ell$-dependent factor is introduced in the parametrization of ${\cal M}_{\mu\gamma^*}$,
\begin{align}
    {\cal M}_{\mu\gamma^*}(k,W,Q^2)=
    R_{\ell'}(\lambda, q/q_\gamma)\bar{\cal M}_{\mu\gamma^*}(k,W,Q^2)\,,
    \label{ampl_2}
\end{align}
with $q_\gamma$ of Eq.~\eqref{eq:qg}; $\lambda$ is a parameter to be fitted, and
\begin{align}
    R_{\ell'}(\lambda,x)={B_{\ell'}(\lambda x)\over B_{\ell'}(\lambda)}
\end{align}
with the Blatt-Weisskopf barrier-penetration factors $B_{\ell'}(r)$~\cite{Blatt:1952,Manley:1984jz} with  the limits
\begin{align}
    B_{\ell'}(r)\sim O(r^{\ell'})
    \;\text{~and~}
    \;B_{\ell'}(r)\sim O(r^0)\,,
    \label{eq:blattweisskopf_limits}
\end{align}
for small and large arguments of $B_{\ell'}(r)$, respectively. The index $\ell'$ relates to $J=\ell\pm$ as
\begin{align}
    \ell' &=
\left\{
\begin{matrix}
\ell~~~~~~~\text{for}~(E_{\ell+},L_{\ell+},L_{1-},M_{\ell+},M_{\ell-}) \ ,\\
\ell-2~~\text{for}~(E_{\ell-},L_{\ell-}\text{  and  }\ell\geq 2)\ ,~~~~~~~
\end{matrix}
\right.
    \label{l_eff}
\end{align}
which ensures that the pseudothreshold constraints tabulated in \eqref{eq:pseudo-thre-conditions} are satisfied. Explicit forms of $B_\ell(r)$, for values of $\ell$ from zero to five, are given in Appendix~\ref{app:blatt-weisskopf}. Note that the transition from small to large arguments in  Eq.~\eqref{eq:blattweisskopf_limits} should happen on a natural scale, i.e., $\lambda$ which is unitless, should be of order one.  The threshold behavior ${\cal M}\sim k^\ell$ tabulated in \eqref{eq:pseudo-thre-conditions} is automatically satisfied by electroproduction multipoles because it is already built into the photoproduction amplitude and this carries over to the virtual-photon case.

To summarize the structure of the fit, each multipole $E_{\mu\gamma^*}$ and $M_{\mu\gamma^*}$ carries $(1+N)$ fit parameters $\beta_0,...,\beta_N$ for the non-pole part $\tilde F^\mu$, plus $(1+N)$ parameters for each $\tilde F_i$ of the $i_\text{max}$ resonances in the pertinent partial wave. The channel-dependence of the non-pole form-factor $\tilde F^\mu (Q^2)$ is not fully used in the current fits, because only pion electroproduction data are analyzed. In $\tilde F^\mu$ we therefore set those $\beta_i$ to zero that couple to the photon but do not correspond to the $\pi N$ channel, i.e., ($\mu\in\{\eta N, K\Lambda,\pi\Delta\}$~\cite{Ronchen:2018ury}).

Furthermore, one could also fit the parameters $Q_w$ and $Q_r$ of Eq.~\eqref{eq:formfactor-FD} to data, but we chose to fix them to the quoted values to avoid over-parametrization. In addition, there is one Blatt-Weisskopf range factor $\lambda$ per multipole. The longitudinal multipoles exhibit the exact same structure of fit parameters as the $E$ and $M$ multipoles, through the functions $D_{\ell\pm}$ of Eq.~\eqref{eq:D-formfactor}. For the two exceptions
($(\ell\pm,I)\in\{(1-,1/2),(1-,3/2)\}$), there are $(1+i_\text{max})$ additional fit parameters $\zeta^{NP}$ for each longitudinal multipole as indicated in Eq.~\eqref{eq:zetanp} for the non-pole part and, similarly, for each of the $i_\text{max}$ resonances. In total, we allow for 209 fit parameters in the parametrization, but we have also explored variants as discussed in Sec.~\ref{subsec:fit scenarios}.

\begin{figure*}[t]
\rule{0.2\linewidth}{0.4pt}
{$\gamma^*p\to\pi^0p$}
\rule{0.2\linewidth}{0.4pt}\\
  \begin{minipage}{0.52\linewidth}
  \begin{center}
    \includegraphics[width=1.1\linewidth,trim=0 0 3.3cm 0,clip]{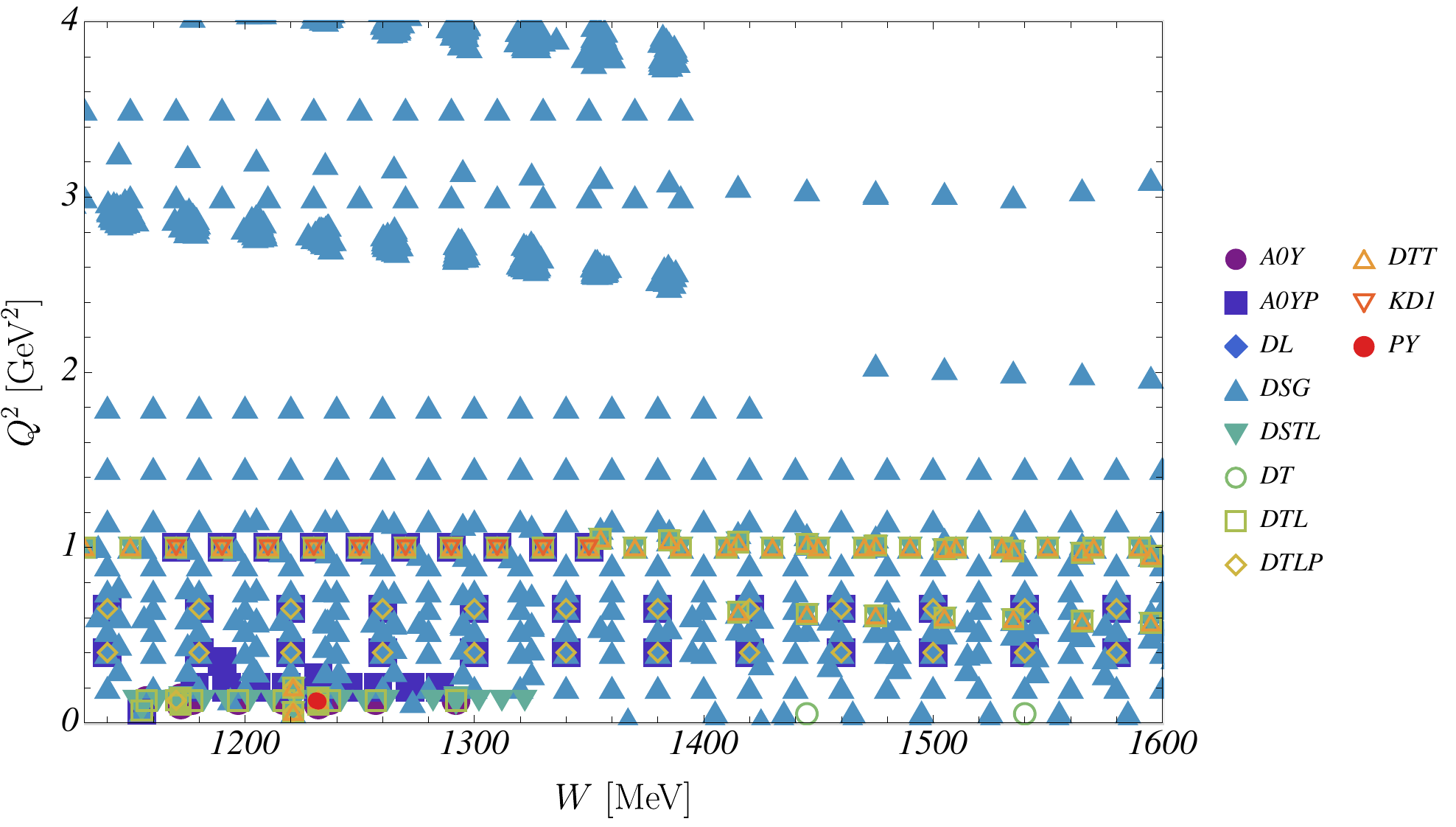}
  \end{center}
  \end{minipage}
  \begin{minipage}{0.47\linewidth}
    \begin{flushleft}
~~~~~
    \scriptsize
    \renewcommand{\arraystretch}{1.62}
    \begin{tabular}{|cl|l|l|}
    \multicolumn{4}{c}{}\\
    \hline
    \multicolumn{2}{|c|}{Type}&$N_{\rm data}$&Ref\\
    \hline
    {\Large\color{Mulberry}$\bullet$}&$\rho_{LT}$&45&\cite{Mertz:1999hp,Elsner:2005cz}\\
    {\color{Blue}$\blacksquare$}&$\rho_{LT'}$&2644&\cite{Joo:2003uc, Sparveris:2002fh, Kelly:2005jj, Bartsch:2001ea,Bensafa:2006wr} \\
    {\color{NavyBlue}$\blacklozenge$}&$\sigma_L$&--&    \\
    {\color{MidnightBlue}$\blacktriangle$}&$d\sigma/d\Omega$&39942&\cite{Laveissiere:2003jf, Ungaro:2006df, Gayler:1971zz, May:1971zza, Hill:1977sy, Joo:2001tw, Frolov:1998pw, Siddle:1971ug, Haidan:1979yqa, Sparveris:2002fh, Kelly:2005jj, Kalleicher:1997qf, Baetzner:1974xy, Latham:1979wea, Latham:1980my, Stave:2006jha, Sparveris:2006uk, Alder:1975xt, Afanasev:1975qa, Shuttleworth:1972nw, Blume:1982uh, Rosenberg:1979zm, Gerhardt:1979zz}\\
    {\color{TealBlue}$\blacktriangledown$}&$\sigma_T+\epsilon \sigma_L$&318&\cite{Laveissiere:2003jf, Mertz:1999hp, Sparveris:2002fh, Kunz:2003we, Stave:2006ea, Sparveris:2006uk, Sparveris:2004jn, Alder:1975xt}\\
    {\Large\color{JungleGreen}$\circ$}&$\sigma_{T}$&10&\cite{Blume:1982uh}    \\
    {\color{LimeGreen}$\square$}&$\sigma_{LT}$&312&\cite{Laveissiere:2003jf, Sparveris:2002fh, Mertz:1999hp, Kunz:2003we, Stave:2006ea, Sparveris:2006uk, Sparveris:2004jn, Alder:1975xt}   \\
    {\color{SpringGreen}$\lozenge$}&$\sigma_{LT'}$&198&\cite{Joo:2003uc, Kunz:2003we, Stave:2006ea, Sparveris:2006uk}    \\
    {\color{Dandelion}$\triangle$}&$\sigma_{TT}$&266&\cite{Laveissiere:2003jf, Stave:2006ea, Sparveris:2006uk, Sparveris:2004jn, Alder:1975xt}    \\
    {\color{RedOrange}$\triangledown$}&$K_{D1}$&1527&\cite{Kelly:2005jj}  \\
    {\Large\color{Red}$\bullet$}&$P_Y$&2&\cite{Warren:1999pq, Pospischil:2000ad}\\
    \hline
    \end{tabular}
    \vspace{1.05cm}
    \end{flushleft}
  \end{minipage}
~\\~\\
\rule{0.2\linewidth}{0.4pt}
{$\gamma^*p\to\pi^+n$}
\rule{0.2\linewidth}{0.4pt}\\
  \begin{minipage}{0.52\linewidth}
  \begin{center}
    \includegraphics[width=1.1\linewidth,trim=0 0 3.3cm 0,clip]{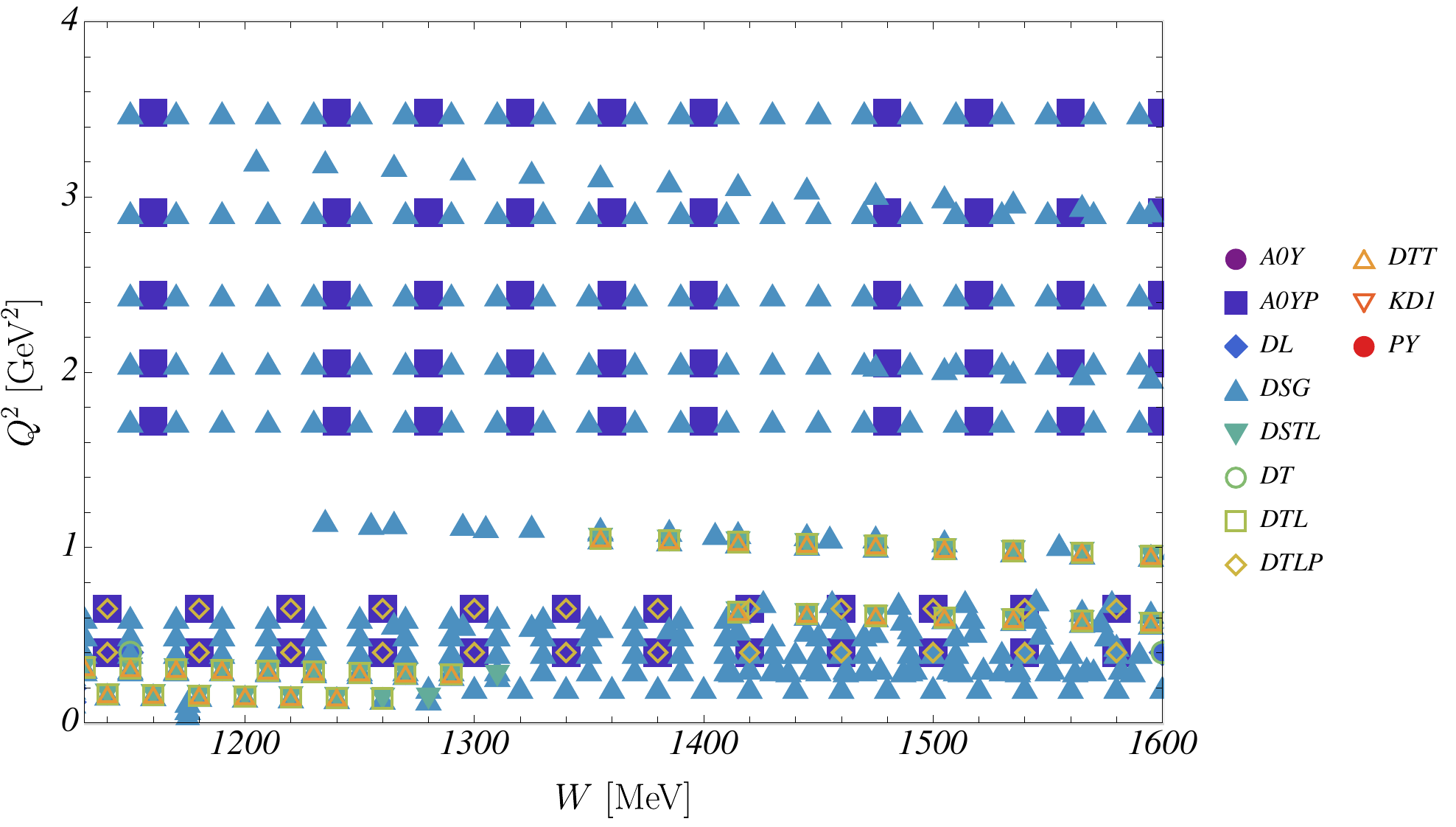}
  \end{center}
  \end{minipage}
  \begin{minipage}{0.47\linewidth}
    \begin{flushleft}
~~~~~
    \scriptsize
    \renewcommand{\arraystretch}{1.62}
    \begin{tabular}{|cl|l|l|}
    \multicolumn{4}{c}{}\\
    \hline
    \multicolumn{2}{|c|}{Type}&$N_{\rm data}$&Ref\\
    \hline
    {\Large\color{Mulberry}$\bullet$}&$\rho_{LT}$&--&--\\
    {\color{Blue}$\blacksquare$}&$\rho_{LT'}$&4354&\cite{Joo:2004mi, PARK-pc-08-2007}   \\
    {\color{NavyBlue}$\blacklozenge$}&$\sigma_L$&2&\cite{Gaskell:2001fn}      \\
    {\color{MidnightBlue}$\blacktriangle$}&$d\sigma/d\Omega$&32813&\cite{Egiyan:2006ks, Breuker:1977vy, PARK-pc-08-2007, Bardin:1975oea, Bardin:1977zu, Gerhardt:1979zz, DAVENPORT-phd1980, Vapenikova:1988fd, Hill:1977sy, Alder:1975na, Evangelides:1973gg, Breuker:1982nw, Breuker:1982um}  \\
    {\color{TealBlue}$\blacktriangledown$}&$\sigma_T+\epsilon \sigma_L$&144&\cite{Breuker:1977vy, Alder:1975na}   \\
    {\Large\color{JungleGreen}$\circ$}&$\sigma_{T}$&2&\cite{Gaskell:2001fn}      \\
    {\color{LimeGreen}$\square$}&$\sigma_{LT}$&106&\cite{Breuker:1977vy, Alder:1975na}    \\
    {\color{SpringGreen}$\lozenge$}&$\sigma_{LT'}$&192&\cite{Joo:2003uc}    \\
    {\color{Dandelion}$\triangle$}&$\sigma_{TT}$&91&\cite{Breuker:1977vy, Alder:1975na}     \\
    {\color{RedOrange}$\triangledown$}&$K_{D1}$&--&--\\
    {\Large\color{Red}$\bullet$}&$P_Y$&--&--\\
    \hline
    \end{tabular}
    \vspace{1.05cm}
    \end{flushleft}
  \end{minipage}
~\\~\\
\caption{
\label{fig:data-coverage}
Overview of the experimental data used in the main fits $(0<Q^2<4~{\rm GeV}^2, 1.13<W<1.6~{\rm GeV})$ for aggregated values of $\theta,\,\phi,\,\epsilon$. Symbol shapes differentiate between different observable types, while the total number of data is 82968. See Sec.~\ref{subsec:observables} and Appendix~\ref{app:sec:Kelly} for the definition of the observables.}
\end{figure*}

\begin{table*}[t]
\renewcommand{\arraystretch}{1.5}
\resizebox{\linewidth}{!}{
\begin{tabular}{|c|cc|cc|cc|cc|cc|cc|cc|cc|cc|cc|cc?c|}
\hline
\multirow{2}{*}{Fit}&
\multicolumn{2}{|c}{$\sigma_L$}&
\multicolumn{2}{|c}{$d\sigma/d\Omega$}&
\multicolumn{2}{|c}{$\sigma_{T}+\epsilon\sigma_{L}$}&
\multicolumn{2}{|c}{$\sigma_{T}$}&
\multicolumn{2}{|c}{$\sigma_{LT}$}&
\multicolumn{2}{|c}{$\sigma_{LT'}$}&
\multicolumn{2}{|c}{$\sigma_{TT}$}&
\multicolumn{2}{|c}{$K_{D1}$}&
\multicolumn{2}{|c}{$P_Y$}&
\multicolumn{2}{|c}{$\rho_{LT}$}&
\multicolumn{2}{|c?}{$\rho_{LT'}$}&
\multirow{2}{*}{$\chi^2_{\rm dof}$}
\\
&
$\pi^0p$&$\pi^+n$&
$\pi^0p$&$\pi^+n$&
$\pi^0p$&$\pi^+n$&
$\pi^0p$&$\pi^+n$&
$\pi^0p$&$\pi^+n$&
$\pi^0p$&$\pi^+n$&
$\pi^0p$&$\pi^+n$&
$\pi^0p$&$\pi^+n$&
$\pi^0p$&$\pi^+n$&
$\pi^0p$&$\pi^+n$&
$\pi^0p$&$\pi^+n$&
\\
\thickhline
$\mathfrak{F}_1$&
 \text{--} & 9 & 65355 & 53229 & 870 & 418 & 87 & 88 & 1212 & 133 & 862 & 762 & 4400 & 251 & 4493 & \text{--} & 234 & \text{--} & 525 & \text{--} & 3300 & 10294 & 1.77 \\
\hline
$\mathfrak{F}_2$&
 \text{--} & 4 & 69472 & 55889 & 1081 & 619 & 65 & 78 & 1780 & 150 & 1225 & 822 & 4274 & 237 & 4518 & \text{--} & 325 & \text{--} & 590 & \text{--} & 3545 & 10629 & 1.69 \\
\hline
$\mathfrak{F}_3$&
\text{--} & 8 & 66981 & 54979 & 568 & 388 & 84 & 95 & 1863 & 181 & 1201 & 437 & 3934 & 339 & 4296 & \text{--} & 686 & \text{--} & 687 & \text{--} & 3556 & 9377 & 1.81 \\
\hline
$\mathfrak{F}_4$&
 \text{--} & 22 & 63113 & 52616 & 562 & 378 & 153 & 107 & 1270 & 146 & 1198 & 1015 & 4385 & 218 & 5929 & \text{--} & 699 & \text{--} & 604 & \text{--} & 3548 & 11028 & 1.78 \\
\hline
$\mathfrak{F}_5$&
 \text{--} & 20 & 65724 & 53340 & 536 & 528 & 125 & 81 & 1507 & 219 & 1075 & 756 & 4134 & 230 & 5236 & \text{--} & 692 & \text{--} & 554 & \text{--} & 3580 & 11254 & 1.81 \\
\hline
$\mathfrak{F}_6$&
 \text{--} & 18 & 71982 & 58434 & 1075 & 501 & 29 & 68 & 1353 & 135 & 1600 & 1810 & 3935 & 291 & 5364 & \text{--} & 421 & \text{--} & 587 & \text{--} & 3932 & 11475 & 1.78 \\
 \hline
\end{tabular}
}
\caption{Fit quality for various fit scenarios described in the main text. All numbers represent $\chi^2$ contributions with respect to different channels and observable types as defined in Eq.~\eqref{eq:chi2}. Total number of data is $82968$ and $91896$ for scenarios 1/3/4/5 and 2/6, respectively. Cases with no data are marked by "--". Rightmost column shows the aggregated $\chi^2_{\rm dof}$.}
\label{tab:fit-results}
\end{table*}

\section{Fits to data}
\label{sec:fits}

\subsection{Database}
\label{subsec:exp-data}

The data used in this study were taken from the extensive SAID
database~\cite{SAID-web}, containing on the order of $10^5$ data for the electroproduction of charged and neutral pions off proton targets. See Fig.~\ref{fig:data-coverage} for an overview of the coverage of the $Q^2$--$W$ plane as well as the number of data points for each observable. More measurements, both in number and type, are available for neutral-pion than for charged pion electroproduction. For both final states, the SAID database is dominated by unpolarized differential cross sections.

Statistically, it is better to fit differential cross sections instead of their components according to Eq.~\eqref{eq:diff-cross-sec.} because on the one hand, it avoids potential bias in the extraction of the components; on the other hand, the data of the components are necessarily correlated but those correlations are typically not quoted in experimental papers. Therefore, we have generally fitted differential cross sections directly, if available, rather than the separated components $\sigma_T$, $\sigma_L$, $\sigma_{TT}$, $\sigma_{LT}$, and $\sigma_{LT'}$. However, to have simpler comparisons with previous analyses, we have included plots of predicted structure function data as well, see Sec.~\ref{subsec:fit results}.

The largest sets of polarized data, included in this fit, involve ratios of structure functions~\eqref{eq:polarization_obs}, and the recoil-polarization measurements of Kelly. The Kelly data~\cite{Kelly:2005jj} give substantial constraints on neutral-pion production at $Q^2$ = 1 ${\rm GeV^2}$. In a recent study~\cite{Tiator:2017cde} of the data types required for model-independent complete-experiment or partial-wave analyses, a subset of the Kelly data was shown to be sufficient for a partial-wave analysis up to P-waves. It should be noted, however, that this study of experimental completeness assumed error-free measurements. In the Kelly paper~\cite{Kelly:2005jy}, fits with different angular-momentum cutoffs were attempted, showing sensitivity to approximations made in their multipole analysis.

Substantial use of SAID and the MAID electroproduction websites~\cite{SAID-web, MAID-web} and codes allowed checks for consistency of conventions, definitions and naming schemes within the SAID database. As described below, a website is being constructed to compare fits by MAID, SAID, and the present analysis to plotted data~\cite{SAID-dev}.

\begin{figure}[b]
    \includegraphics[width=\linewidth]{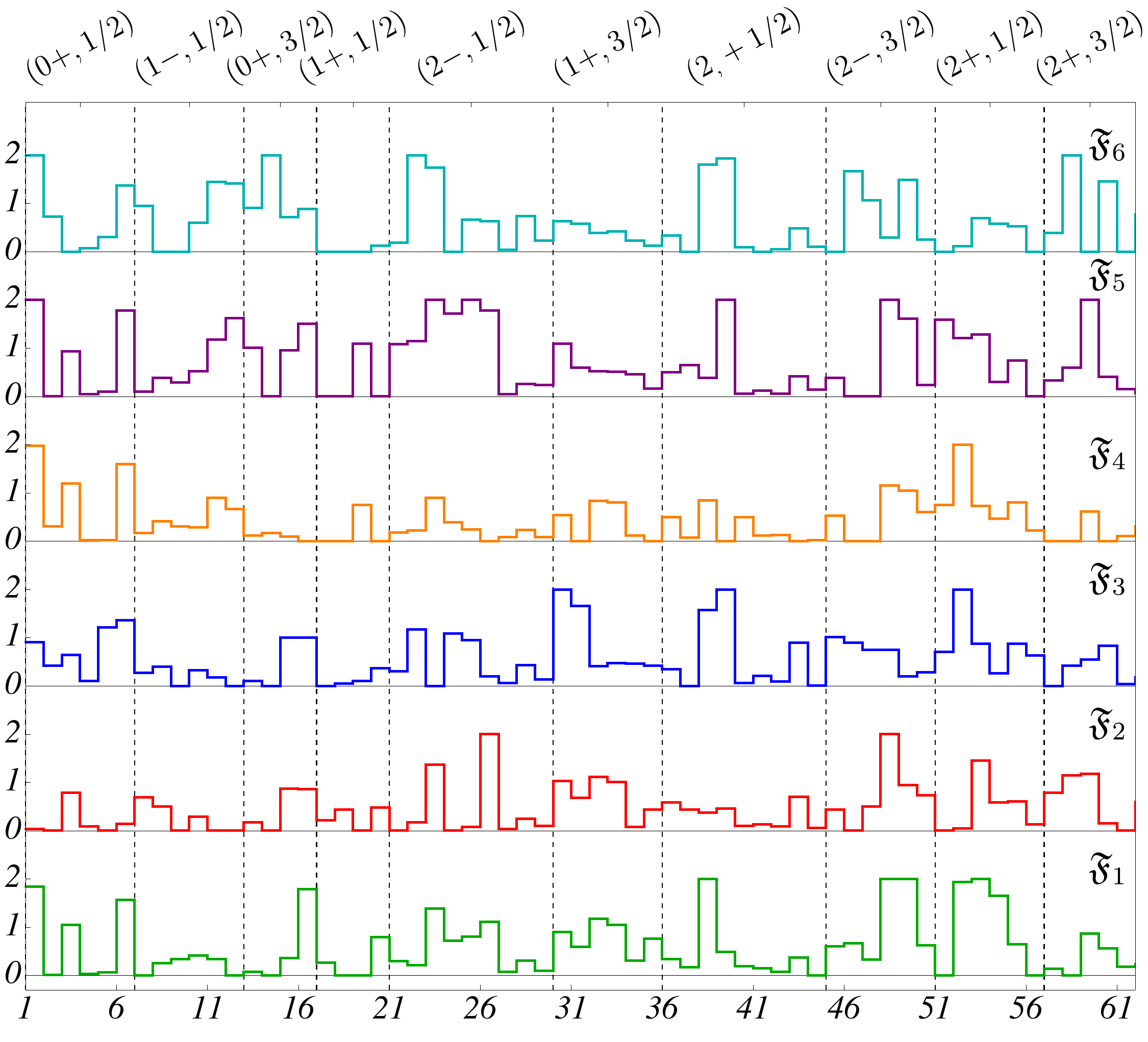}
\caption{
Values of 62 $\beta_0$ parameters with respect to different fit scenarios ($\mathfrak{F}_i$) and partial wave $(\ell\pm,I)$ where $I$ denotes isospin.}
\label{fig:beta0-cut}
\end{figure}

\begin{figure*}[t]
    \includegraphics[width=\linewidth, trim=0 0 0.1cm 0, clip]{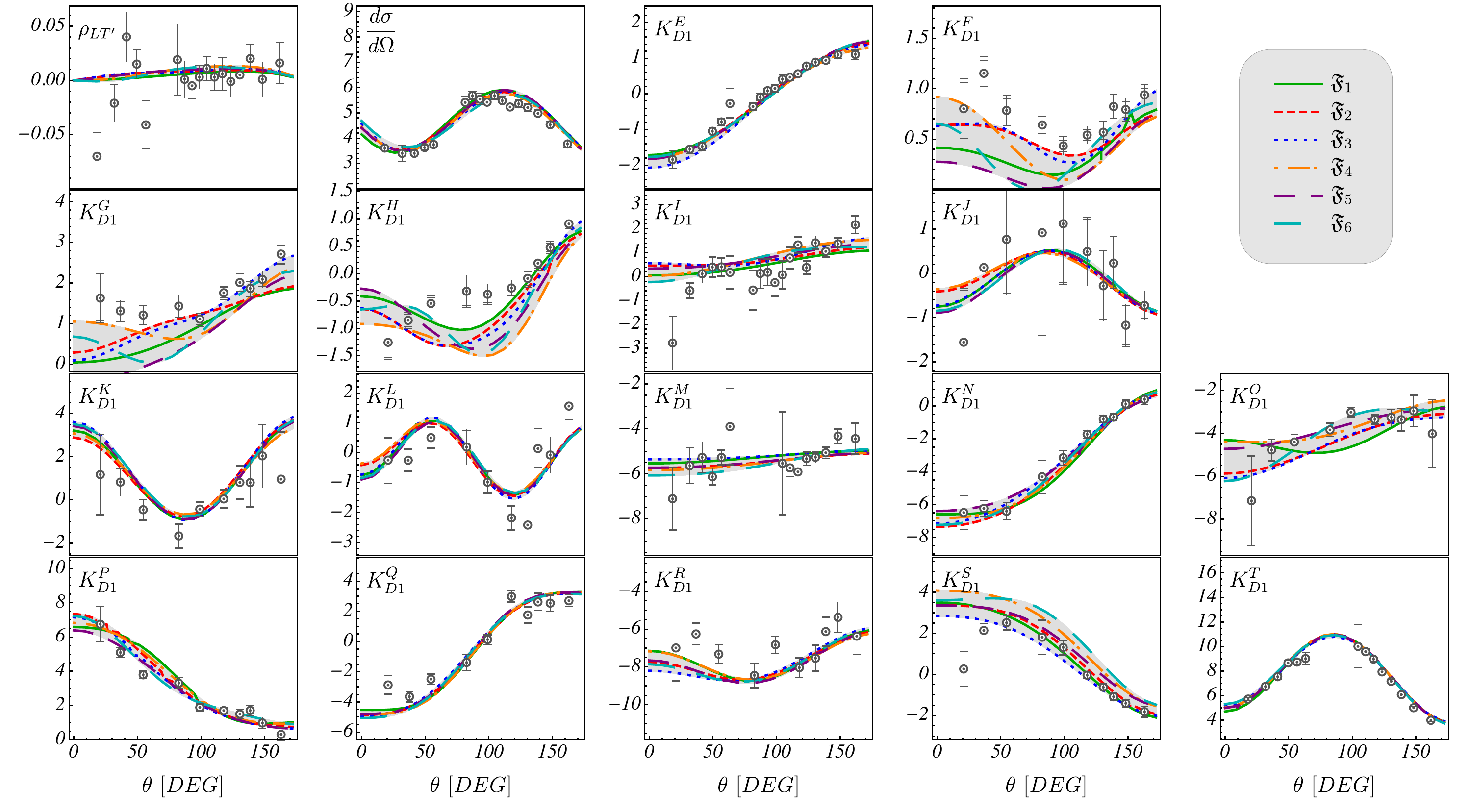}
\caption{Comparison of the best fits to the Kelly data~\cite{Kelly:2005jj} (open circles with error bars) at $Q^2=1~{\rm GeV}^2$, $W=1.23$~GeV, $\phi=15^\circ$ in the $\pi^0p$ channel. Different curves correspond to various fit strategies, representing systematic uncertainty of our approach. Note that $\rho_{LT'}$ is unitless, while others are given in $[\rm \mu b/sr]$. In this and subsequent figures, the shading between the curves is included to guide the eye.}
\label{fig:Kelly05-1230}
\end{figure*}

\begin{figure*}[t]
    \includegraphics[width=1.0\linewidth, trim= 0 1.1cm 0 0, clip]{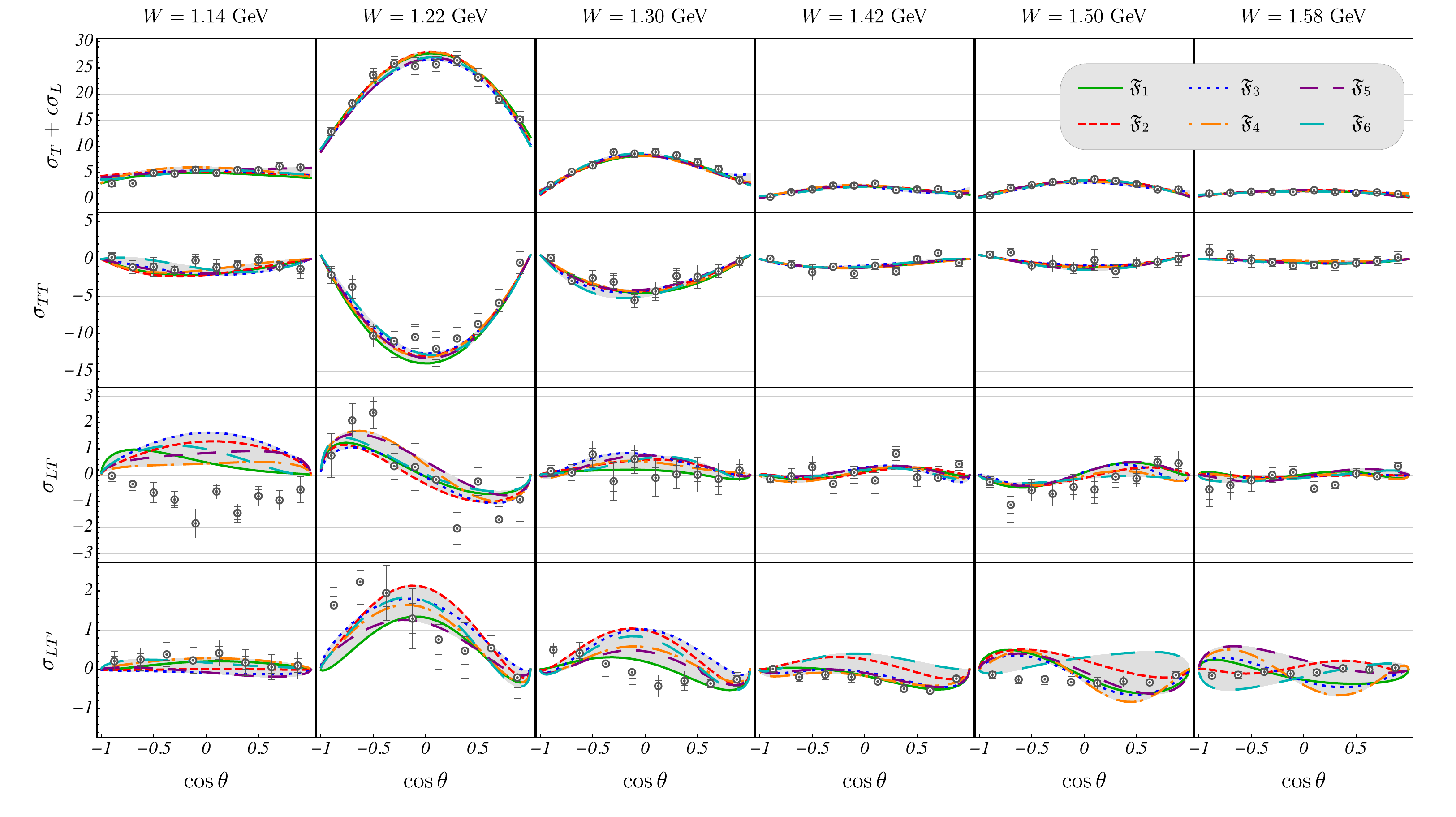}
    \includegraphics[width=1.0\linewidth, trim= 0 0 0 0.8cm, clip]{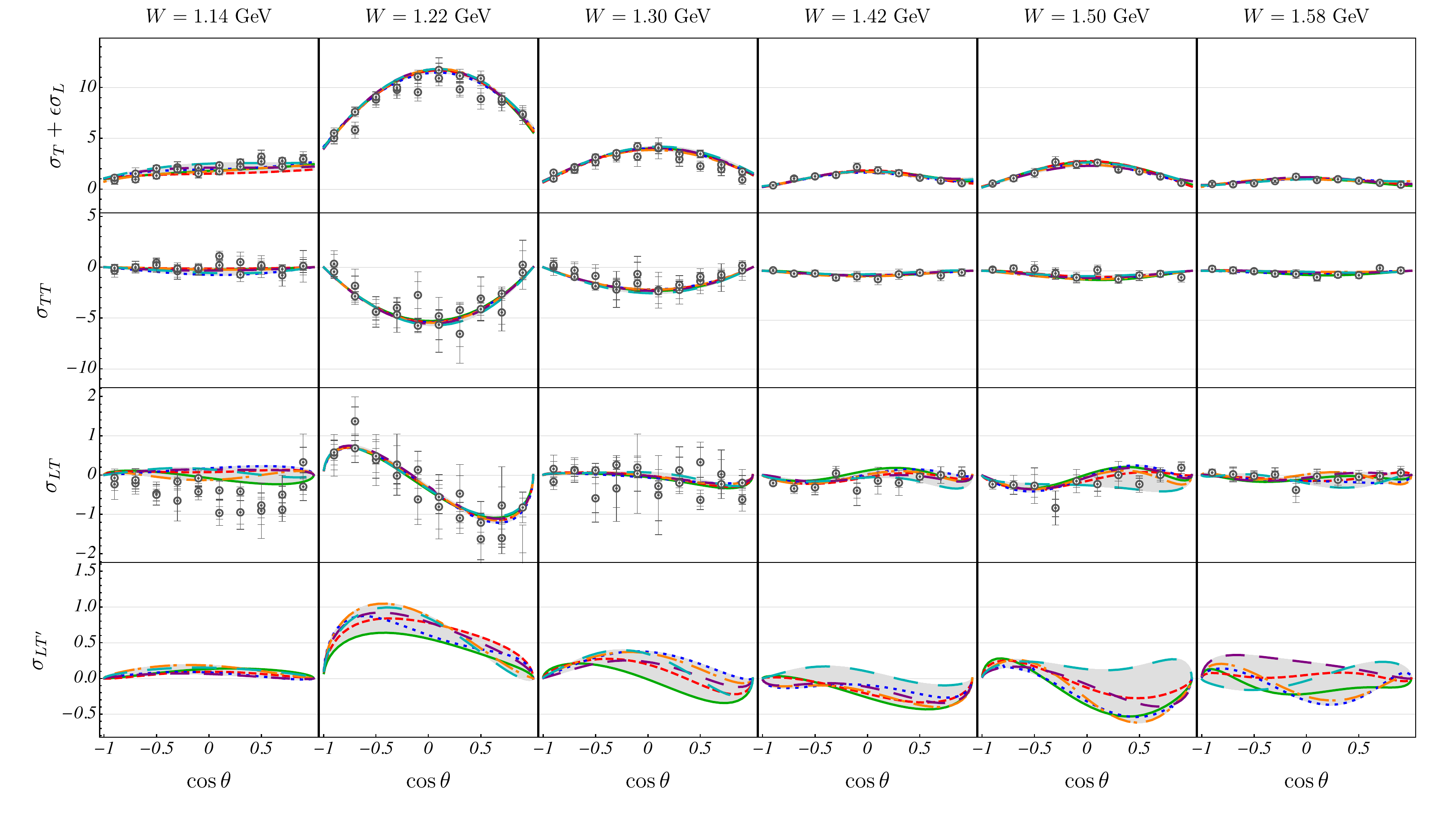}
\caption{
Best fits to electroproduction data for $Q^2=0.4~{\rm GeV}^2$ (top) and $Q^2=0.9~{\rm GeV}^2$ (bottom) compared to the structure functions from Refs.~\cite{Joo:2001tw,Joo:2003uc}, shown by open circles with statistical and sum of statistical and systematic error bars, respectively. All units are $[\rm \mu b/sr]$. Different curves correspond to various fit strategies, representing systematic uncertainty of our approach.
}
\label{fig:JULIA}
\end{figure*}

\subsection{Fit scenarios }
\label{subsec:fit scenarios}

The parametrization of the multipoles introduced in Sec.~\ref{subsec:JuBo extended} is subject to a large set of free parameters. In particular, considering S-, P- and D-waves and both isospin-channels $I=(1/2,3/2)$ leads to 62 new parameters for each new order in the expansion of $P^N$ used to parametrize $E$, $M$ and $L$ multipoles. In this context, our preliminary fits have shown that $N=2$ (see Eqs.~\eqref{eq:formfactor-Ftilde} and \eqref{eq:D-formfactor}) yields a sufficiently flexible parametrization, without clear over-fitting of data. See discussion at the end of Sec.~\ref{subsec:JuBo extended} for additional details. More extensive statistical studies are beyond  the scope of the present paper and will be discussed in a future work.

In addition to the $\beta$-type parameters, pseudo-threshold regulating parameters ($\lambda$)  and normalization factors ($\zeta$) for two longitudinal multipoles ($L_{1-,1/2}$ and $L_{1-,3/2}$) not fixed by Siegert's conditions add 18 and 5 new parameters, respectively. Thus, the total number of parameters sums up to 209.

The free parameters are fit to reproduce the database described in Sec.~\ref{subsec:exp-data} by minimizing the $\chi^2$ function
\begin{align}
    \chi^2=\sum_{i=1}^{N_{\rm data}}\left(\frac{\mathcal{O}_i^{\rm exp}-\mathcal{O}_i}{\Delta_i^{\rm stat}+\Delta^{\rm syst}_i}\right)^2\,.
    \label{eq:chi2}
\end{align}
To report the results we also define $\chi^2_{\rm dof}=\chi^2/(N_{\rm data}-209)$. We note that inclusion of systematic errors can be done at different levels of rigor. For example, the SAID group allows data to be ``floated'' with a $\chi^2$ penalty determined by the overall systematic error~\cite{Doring:2016snk}. The J\"uBo/GW group used  similar normalization freedom in some of the more recently included data sets~\cite{Collins:2017sgu,Strauch:2015zob}. In this study, we add the systematic error to the statistical one as indicated in Eq.~\eqref{eq:chi2} following a procedure widely used in the field; effectively, one neglects the correlations between data due to systematic effects. As there is no reason to believe that systematic effects should be Gaussian, we simply add the uncertainties linearly. In some baryon resonance analyses data are weighted with factors enhancing the influence of sparsely measured observables. In the present study, we did not follow this procedure.

Given the high dimensionality of the minimization problem, an obvious question arises about the statistical significance of the present solution. To our knowledge, a systematic way to answer this question does not exist. Thus, to get an understanding of the $\chi^2$ landscape we have performed a series of fits --- denoted by $\mathfrak{F}_i$ with respect to kinematic ranges, and strategy of a minimization. For example, in $\mathfrak{F}_1$ vs.\ $\mathfrak{F}_3$ vs.\ $\mathfrak{F}_4$, we have studied the importance of the choice of starting values by performing preliminary fits to various subsets of data, i.e., 1/32, 1/128 and complete set, respectively. In $\{\mathfrak{F}_1,..,\mathfrak{F}_4\}$ vs.\ $\{\mathfrak{F}_5,\mathfrak{F}_6\}$ we have changed the fitting strategy from sequential fits (increasing step-wise the number of free parameters from 32 to 209 by including higher and higher partial waves) to simultaneous fits with 209 parameters, all of them set to zero initially. Finally, we have also checked in $\mathfrak{F}_1$ vs.\ $\mathfrak{F}_2$ and in $\mathfrak{F}_5$ vs.\ $\mathfrak{F}_6$ the stability of the results when adding more data by increasing the limit $Q^2<4~{\rm GeV}\to Q^2<6~{\rm GeV}$ corresponding to $N_{\rm data}=82968\to N_{\rm data}=91896$. We expect that systematic effects associated with these strategies significantly exceed the statistical uncertainties.

\begin{figure*}
    \includegraphics[width=\linewidth,trim=0 0 0 0]{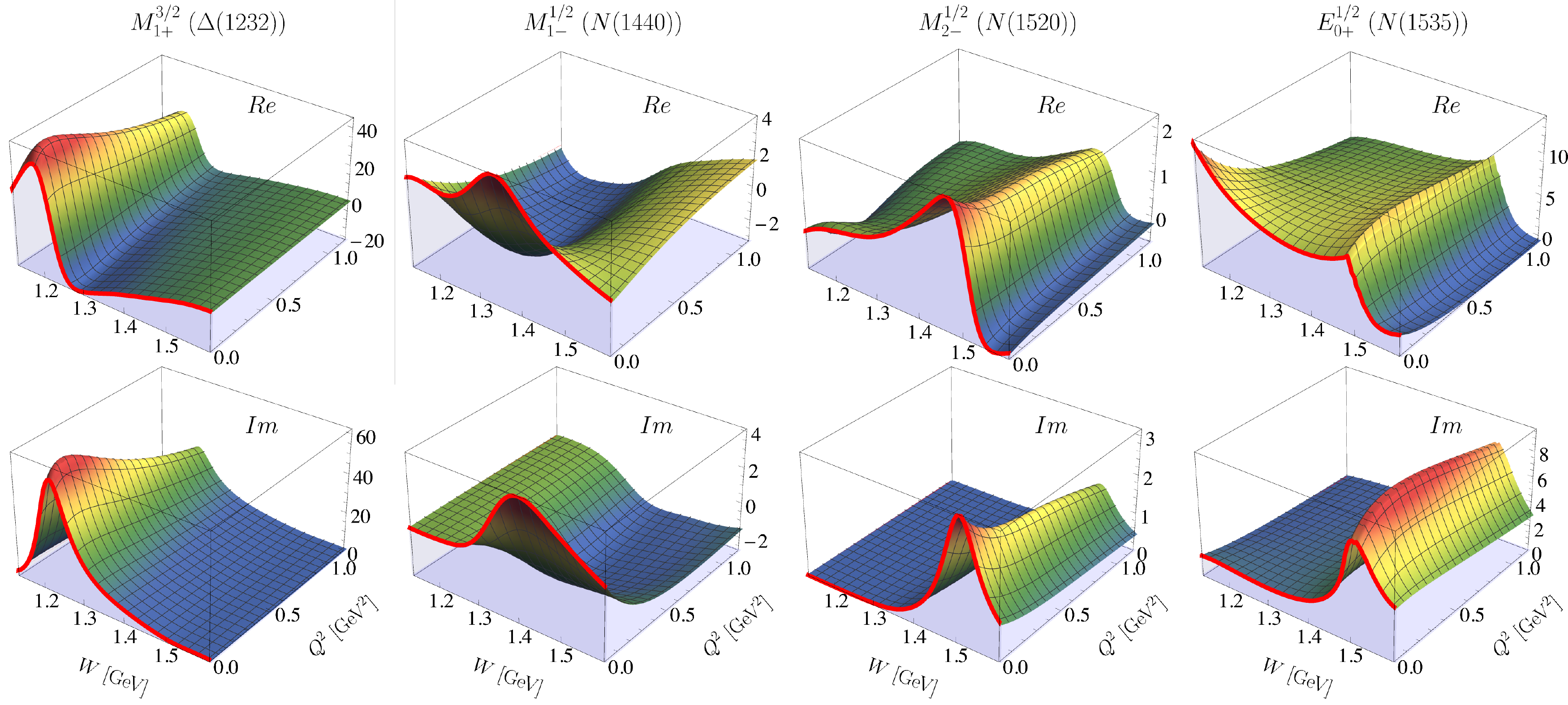}
\caption{
Selected results (representative fit $\mathfrak{F}_1$) for multipoles associated with $\Delta(1232)$, $N(1440)$, $N(1520)$ and $N(1535)$, respectively. At $Q^2=0~{\rm GeV}^2$ the solution is constrained by pion-induced and photoproduction data via the J\"ulich-Bonn model (red line), while the extension into $Q^2>0$ is facilitated by the current parametrization. All in units of mfm.
}
\label{fig:Roper-and-friends}
\end{figure*}

\subsection{Fit results}
\label{subsec:fit results}

The quality of the obtained best fits $\mathfrak{F}_i$ is recorded in Table~\ref{tab:fit-results}. There we also record contributions from each observable type to the total $\chi^2$. We note that the best $\chi^2$ is found to be consistent among all fit strategies, which indicates the overall flexibility of the parametrization. As to the question whether the found minima are identical, the 209-dimensional parameter space is unwieldy to use in addressing this question. Instead, a cut through the 62 dimensions of the crucial $\beta_0$ parameters, shown in Fig.~\ref{fig:beta0-cut}, suggests no clear similarity between the best-fit parameters of individual fits. While any conclusion drawn from this observation is prone to possibly large correlations between different parameters, it seems that the solutions shown in Table~\ref{tab:fit-results} represent different local minima of the $\chi^2$ landscape. This strengthens our previous assumption that including various fit scenarios is indeed a fair representation of the uncertainties.

The fit results suggest that the approach is sufficiently flexible to provide a good overall fit to the data. Going beyond arguments based on the overall $\chi^2$, we display the fit quality for selected observables. In particular, a large set of differential cross section and recoil polarization data is provided by Kelly~\cite{Kelly:2005jj}, see Sec.~\ref{subsec:exp-data}. Fig.~\ref{fig:Kelly05-1230} shows the quality of fit to the Kelly data for a representative kinematical configuration. Except for a few points, the data are described very well. While the Kelly set has a higher $\chi^2$ per datum than displayed in Table I for the full database, i.e. a $\chi^2 \approx 3$ per data point, approximately 1/3 of the $\chi^2$ comes from 1\% outliers. Similar characteristics are found in other observables. In cases where the outliers could be attributed to simple errors in digitizing older data sets, they were not included in our fits.

Another large set of data, taken in a dedicated Jefferson Lab experiment~\cite{Joo:2001tw,Joo:2003uc} was analyzed by EBAC (Excited Baryon Analysis Center) in Ref.~\cite{JuliaDiaz:2009ww}. The data considered in the EBAC analysis was composed of structure functions $\{\sigma_{T}+\epsilon\sigma_{L}, \sigma_{TT}, \sigma_{LT}, \sigma_{LT'}\}$, in contrast to the present fit, where we included the data of Refs.~\cite{Joo:2001tw,Joo:2003uc} via $d\sigma/d\Omega$, $A_{0Y'}$ and $\sigma_{LT'}$ observables. Note also, that Ref.~\cite{JuliaDiaz:2009ww} covered a smaller kinematical range, but the model used there had fewer free parameters. The present fits versus structure functions are shown in Fig.~\ref{fig:JULIA} for $Q^2=0.4~{\rm GeV}^2$ and $0.9~{\rm GeV}^2$. We note that the description of the structure functions does indeed agree with the results of Refs.~\cite{Joo:2001tw,Joo:2003uc} in nearly all cases, apart from a discrepancy in $\sigma_{LT}$ at $W=1.14$~GeV. This energy lies only 10~MeV above the lower limit of data considered in this study and structure functions were not used in the fits directly.

\begin{figure}[b]
    \includegraphics[width=\linewidth,trim=0.1cm 0 2cm 0, clip]{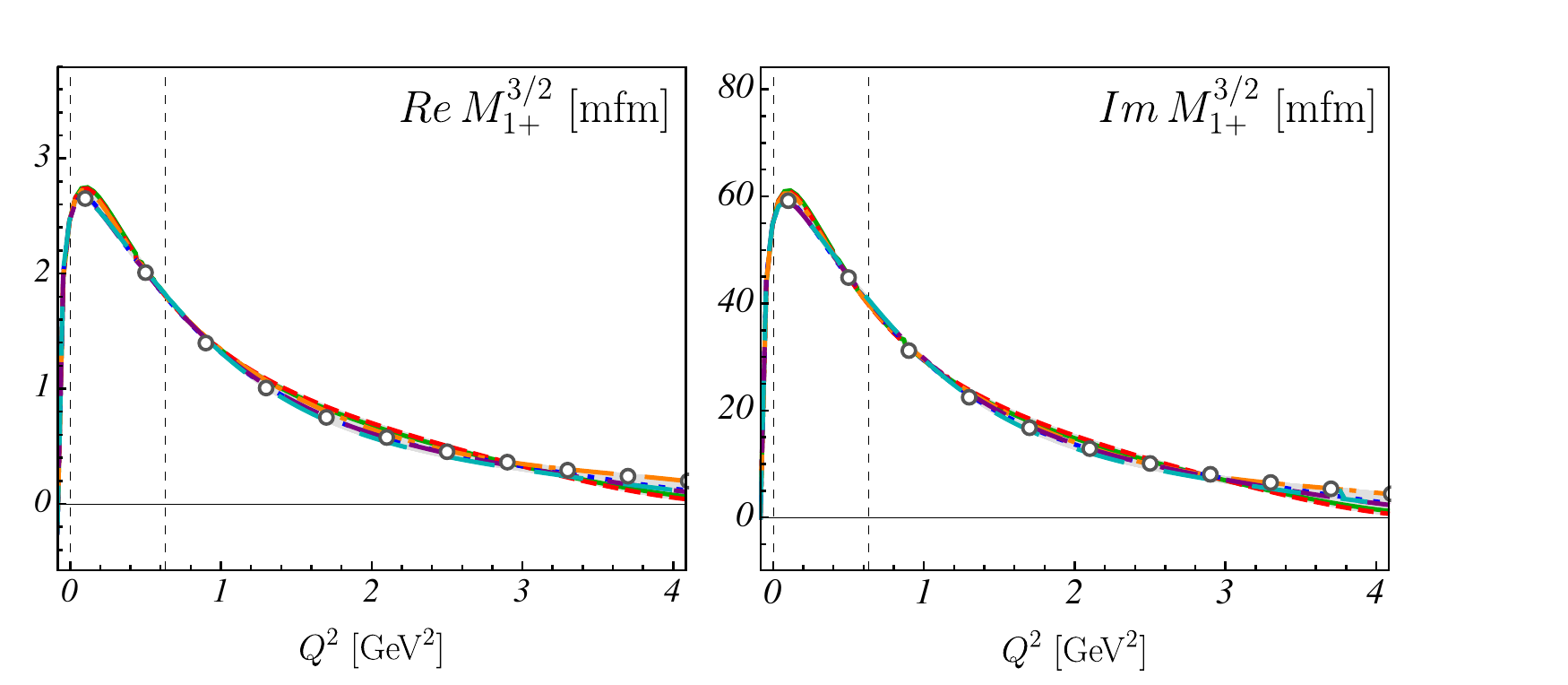}
\caption{
$M_{1+}^{3/2}$ multipole at $W=1230$~MeV for six best fits (color coding as in Fig.~\ref{fig:Kelly05-1230}). Results of MAID2007 analysis~\cite{Drechsel:2007if} are given by open circles for comparison only. Dashed vertical lines show pseudo-threshold, photo-point and $Q^2(\omega=0)$, respectively.
}
\label{fig:M-Wfix@1230}
\end{figure}

With the sets of model parameters determined, we are now able to plot the $E$, $L$ and $M$ multipoles as a function of $Q^2$ and $W$. This is demonstrated in a set of plots for some fixed typical values of $Q^2$ and $W$. In particular, in Figs.~\ref{fig-app:ELM-W=1230MeV},~\ref{fig-app:ELM-W=1380MeV} and~\ref{fig-app:ELM-W=1535MeV} of Appendix~\ref{app:multipoles-W-fixed} we have evaluated the obtained multipoles at fixed energies corresponding to either the Breit-Wigner mass or real part of the pole position associated with the $\Delta (1232)$, $N(1440)$, and $N(1535)$; $W=$1230 MeV, 1380 MeV, and 1535 MeV respectively.
We observe that in many cases the otherwise very weakly constrained longitudinal multipoles are indeed restricted by coupling them to the electric ones at the pseudothreshold point.
Then, for two values $Q^2=0.2~{\rm GeV}^2$ and $1.0~{\rm GeV}^2$, we demonstrate the full set of multipoles, up to $J=2+$, in Figs.~\ref{fig-app:ELM-Q2=0.2} and~\ref{fig-app:ELM-Q2=1}, respectively. Results of the MAID2007 analysis~\cite{Drechsel:2007if} are displayed there for comparison (open circles refer to the MAID2007 energy-dependent solution evaluated at a discrete set of $W$ and $Q^2$).
Note that also near the photon point the MAID2007~\cite{Drechsel:2007if} and J\"uBo~\cite{Ronchen:2018ury}, analyses show sizable differences in some photoproduction multipoles.

Examples of  multipoles dominated by nucleon resonances are depicted in Fig.~\ref{fig:Roper-and-friends}. The $\Delta (1232)$ and $N(1520)$ display canonical resonance behavior while the $\eta$ threshold cusp is evident for the $N(1535)$. While the $\Delta(1232)$ shape disappears at higher $Q^2$, the $N(1535)$ disappears slower. The `profile' of the enigmatic Roper resonance, $N(1440)$, shows a nontrivial $Q^2$ behavior, including zeroes for real and imaginary parts. A better quantitative understanding of the $Q^2$ dependence of resonance will be facilitated by an upcoming analysis at the resonance poles. But it is already reassuring that, e.g., the multipole of the Roper resonance exhibits zeros similar to the pertinent resonance helicity coupling from experiment~\cite{Aznauryan:2004jd, Aznauryan:2009mx, Dugger:2009pn, Mokeev:2013kka, Burkert:2017djo} that is predicted by theoretical approaches~\cite{Segovia:2015hra}.

An extended plot of the $M_{1+}^{3/2}$ multipole is given in Fig.~\ref{fig:M-Wfix@1230} with a comparison to the MAID values. All fits agree with rather small uncertainties. Note that the first vertical line corresponds to a pseudo-threshold point, $Q_{{\rm PT}-}$, at which condition~\eqref{eq:pseudo-thre-conditions} and Siegert's  condition~\eqref{eq:Siegerts_condition} are implemented by construction, whereas the results of MAID2007~\cite{Drechsel:2007if} are restricted to $Q^2\ge 0$. While we expect the very large imaginary part of the multipole to be similar in all fits, including MAID, we also observe close agreement for the real part.

A direct comparison with the EBAC fit~\cite{JuliaDiaz:2009ww} is not possible but the variations in their single-$Q^2$ fits emphasize the lack of sufficient data constraints in the low-$Q^2$ region, and a benefit from including the $Q^2 = 0$ values obtained in photoproduction analyses.

\section{Conclusions and outlook}
\label{sec:conclusions}

In the current paper, we have introduced a novel phenomenological parametrization of the meson electroproduction multipoles. This parametrization builds upon the latest J\"ulich-Bonn solution which includes information from a large database on meson-baryon scattering and meson photoproduction, and takes into account constraints from unitarity, analyticity and chiral symmetry. This approach is extended to the electroproduction sector by parametrizing the $Q^2$ dependence with a general analytic form incorporating constraints from Siegert's theorem and (pseudo)threshold behavior of production amplitudes. Additionally, form factors are included to ensure the fall-off of multipoles at large $Q^2$.

Overall, multipoles of up to D-waves are fitted with respect to 209 parameters to reproduce world pion electroproduction data over a large range of $1.13~{\rm GeV}<W<1.6~{\rm GeV}$ and $Q^2<6~{\rm GeV}^2$. This range is similar to that of Ref.~\cite{Kamano:2018sfb} but we include more polarization observables. We found a good description ($\chi^2_{\rm dof}=1.69-1.81$) of an extensive database ($\sim10^5$ data) including observables of polarized and unpolarized types. Additionally, we have provided an uncertainty estimate on obtained multipoles by exploring several fitting scenarios. Taking these uncertainties into account, the predicted multipoles agree qualitatively with those of the previous MAID2007 analysis~\cite{Drechsel:2007if}.
This is a non-trivial result keeping in mind the large variety of  fit strategies employed in our study, with substantially different initial parameter sets.

In parallel to the present study, a website has been developed~\cite{SAID-dev} utilizing an SQL database, a more modern web framework (Django), and an interactive graphics front end (Plotly), serving as a platform to compare different analyses against the existing data. Presently included models are from the J\"ulich-Bonn fits (J\"uBo2017), describing photoproduction of
pions, kaons, and etas, as well as MAID2007~\cite{Drechsel:2007if},
ETA-MAID~\cite{Chiang:2001as,Tiator:2018heh}, and KAON-MAID~\cite{Bennhold:1999mt}
for photo- and electroproduction of pions, etas, and kaons.
A present version of our electroproduction fits is currently being incorporated.

We plan to extend the formalism to include $\eta N$ and $K\Lambda$ final states. While technically straightforward, the further expansion of the parameter space may need to be reassessed using, e.g., model selection techniques as done in Ref.~\cite{Landay:2018wgf}. Furthermore, a combined fit of pion-induced, photo- and electroproduction data is planned, which will allow for a more reliable extraction of the helicity couplings of resonances.


\begin{center}
{\bf Acknowledgements}
\end{center}
This work is supported by the U.S. Department of Energy grants DE-SC0016582 and DE-SC0016583, and DOE Office of Science, Office of Nuclear Physics under contract DE-AC05-06OR23177. It is also supported by the NSFC and the Deutsche Forschungsgemeinschaft (DFG, German Research
Foundation) through the funds provided to the Sino-German Collaborative Research Center TRR110 “Symmetries and the Emergence of Structure in QCD” (NSFC Grant No. 12070131001, DFG Project-ID 196253076 - TRR 110). The multipole calculation and parameter optimization are performed on the Colonial One computer cluster~\cite{Colonial-One}. The authors gratefully acknowledge the computing time granted through JARA on the supercomputer JURECA~\cite{jureca} at Forschungszentrum Jülich that was used to produce the input at the photo-point.

\bibliography{BIB3.bib,NON-INSPIRE.bib}

\clearpage
\begin{onecolumngrid}
\appendix

\section{Appendix}

\subsection{Definition of $K_{1D}$ observables}
\label{app:sec:Kelly}

\begin{table}[h]
\centering
\begin{tabular}{|r|c|c|l|}
\hline
& SAID/current notation &  Kelly Notation  & Expression in Helicity amplitudes ($H_i$)\\
\hline
1  & $K_{D1}^A$  & RL       &$ (H_5^2 +H_6^2 )/\xi^2$\\
2  & $K_{D1}^B$  & RL(n)    &$ \Im( -H_6H_5^* )2/(\xi^2\sin\theta)$\\
3  & $K_{D1}^C$  & RT       &$ (H_1^2+H_2^2+H_3^2+H_4^2)/2 $\\
4  & $K_{D1}^D$  & RT(n)    &$ \Im( +H_3H_1^* +H_4H_2^* )/\sin\theta$\\
5  & $K_{D1}^E$  & RLT      &$ \Re( +H_5H_1^* -H_5H_4^* +H_6H_2^* +H_6H_3^* )/(\sqrt{2}\xi \sin\theta)$\\
6  & $K_{D1}^F$  & RLT(n)   &$ \Im( -H_2H_5^* -H_3H_5^* +H_1H_6^* -H_4H_6^* )/(\sqrt{2}\xi) $\\
7  & $K_{D1}^G$  & RLT(l)   &$ \Im( +H_1H_5^* +H_4H_5^* +H_2H_6^* -H_3H_6^* )/(\sqrt{2}\xi \sin\theta)$\\
8  & $K_{D1}^H$  &   RLT(t) &$ \Im( +H_2H_5^* -H_3H_5^* -H_1H_6^* -H_4H_6^* )/(\sqrt{2}\xi) $\\
9  & $K_{D1}^I$  &  RLT(h)  &$ \Im( -H_1H_5^* +H_4H_5^* -H_2H_6^* -H_3H_6^* )/(\sqrt{2}\xi \sin\theta)$\\
10 & $K_{D1}^J$  &  RLT(hn) &$ \Re( -H_5H_2^* -H_5H_3^* +H_6H_1^* -H_6H_4^* )/(\sqrt{2} \xi)$\\
11 & $K_{D1}^K$  &  RLT(hl) &$ \Re( -H_5H_1^* -H_5H_4^* -H_6H_2^* +H_6H_3^* )/(\sqrt{2}\xi \sin\theta)$\\
12 & $K_{D1}^L$  & RLT(ht)  &$ \Re( -H_5H_2^* +H_5H_3^* +H_6H_1^* +H_6H_4^* )/(\sqrt{2}\xi )$\\
13 & $K_{D1}^M$  & RTT      &$ \Re( -H_1H_4^* +H_2H_3^* )/( \sin\theta )^2 $\\
14 & $K_{D1}^N$  & RTT(n)   &$ \Im( -H_2H_1^* -H_4H_3^* )/\sin\theta$\\
15 & $K_{D1}^O$  & RTT(l)   &$ \Im( +H_4H_1^* -H_3H_2^* ) /( \sin\theta )^2$\\
16 & $K_{D1}^P$  & RTT(t)   &$ \Im( +H_2H_1^* -H_4H_3^* )/\sin\theta$\\
17 & $K_{D1}^Q$  & RTT(hl)  &$ \Re( -H_1H_1^* -H_2H_2^* +H_3H_3^* +H_4H_4^*)/2 $\\
18 & $K_{D1}^R$  & RTT(ht)  &$ \Re( +H_1H_3^* +H_2H_4^* )/\sin\theta$\\
19 & $K_{D1}^S$  & RL+RT(n)     &$ (H_5^2+H_6^2)/\xi^2 - \Im(H_3^*H_1+H_4^*H_2)/\sin(\theta)$\\
20 & $K_{D1}^T$  & RL+RT      &$ (H_1^2+H_2^2+H_3^2+H_4^2)/2 + (H_5^2+H_6^2)/\xi^2$\\
\hline
\end{tabular}
\caption{\label{tab:KE05-conversion}
Conversion of Kelly observables~\cite{Kelly:2005jj} to the SAID~\cite{SAID-web} notation. Last column shows the implementation in terms of the Helicity amplitudes as employed in this work, while $\xi=\omega/\sqrt{Q^2}$.}
\end{table}

\subsection{Photoproduction kernel in the J\"uBo model}
\label{app:sec:photo_details}
In the J\"uBo approach the photoproduction kernel $V_{\mu\gamma}$ in Eq.~(\ref{eq:jubovg}) is parameterized by the quantities $\gamma^c_{\gamma;i}$ and $\alpha^{\text{NP}}$ which are constructed with energy-dependent polynomials $P^P$ and $P^{\text{NP}}$~\cite{Ronchen:2014cna},
\begin{eqnarray}
\alpha^{\text{NP}}_{\mu\gamma}(p,W)= \frac{ \tilde{\gamma}^a_{\mu}(p)}{\sqrt{m}} P^{\text{NP}}_\mu(W)  \quad\text{and}
\quad
\gamma^c_{\gamma;i}(W)= \sqrt{m} P^{\text P}_i(W)\,.
\label{eq:vg_poly}
\end{eqnarray}
The vertex function $\tilde{\gamma}^a_{\mu}$ is equal to $\gamma^a_{\mu;i}$ but independent of the resonance number $i$. The polynomials $P$ are explicitly given by
\begin{eqnarray}
P^{\text P}_i(W)&=&  \sum_{j=1}^{l_i}  g^{\text P}_{i,j} \left( \frac{W - E_s}{m} \right)^j e^{-\lambda^{\text P}_{i}(W-E_s)}\,,
 \\
P^{\text{NP}}_\mu(W) &=&  \sum_{j=0}^{l_{\mu}} g^{\text{NP}}_{\mu,j} \left( \frac{W - W_s}{m}\right)^j
 e^{-\lambda^{\text{NP}}_{\mu}(W-W_s)}\,.
\label{eq:polys}
\end{eqnarray}
Here, $g^{{\text{P}}({\text{NP}})}$ and $\lambda^{{\text{P}}({\text{NP}})}> 0$ are multipole-dependent free parameters that are fitted to data. The upper limits of the summation $l_i$ and $l_\mu$ are chosen as demanded by the data. In Ref.~\cite{Ronchen:2018ury}, which is used as input for the present study, $l_i$, $l_\mu\le 3$ is sufficient to achieve a good fit result. The expansion point $E_s$ is chosen as $W_s=1077$~MeV in order to be close to the $\pi N$ threshold.

\subsection{Blatt-Weisskopf barrier-penetration factors}
\label{app:blatt-weisskopf}
For $\ell=0,...,5$ the Blatt-Weisskopf barrier-penetration factors~\cite{Blatt:1952,Manley:1984jz} are explicitly given by
\begin{eqnarray}
B_0(r)&=&1\,,\nonumber \\
B_1(r)&=&r/\sqrt{1+r^2}\,,\nonumber\\
B_2(r)&=&r^2/\sqrt{9+3r^2+r^4}\,,\nonumber\\
B_3(r)&=&r^3/\sqrt{225+45r^2+6r^4+r^6}\,,\nonumber\\
B_4(r)&=&r^4/\sqrt{11025+1575r^2+135r^4+10r^6+r^8}\,,\nonumber\\
B_5(r)&=&r^5/\sqrt{893025+99225r^2+6300r^4+315r^6+15r^8+r^{10}}\,.
\label{BW_fctr}
\end{eqnarray}

\clearpage
\subsection{Multipoles for fixed $W$}
\label{app:multipoles-W-fixed}

\begin{figure*}[h!]
    \includegraphics[width=\linewidth]{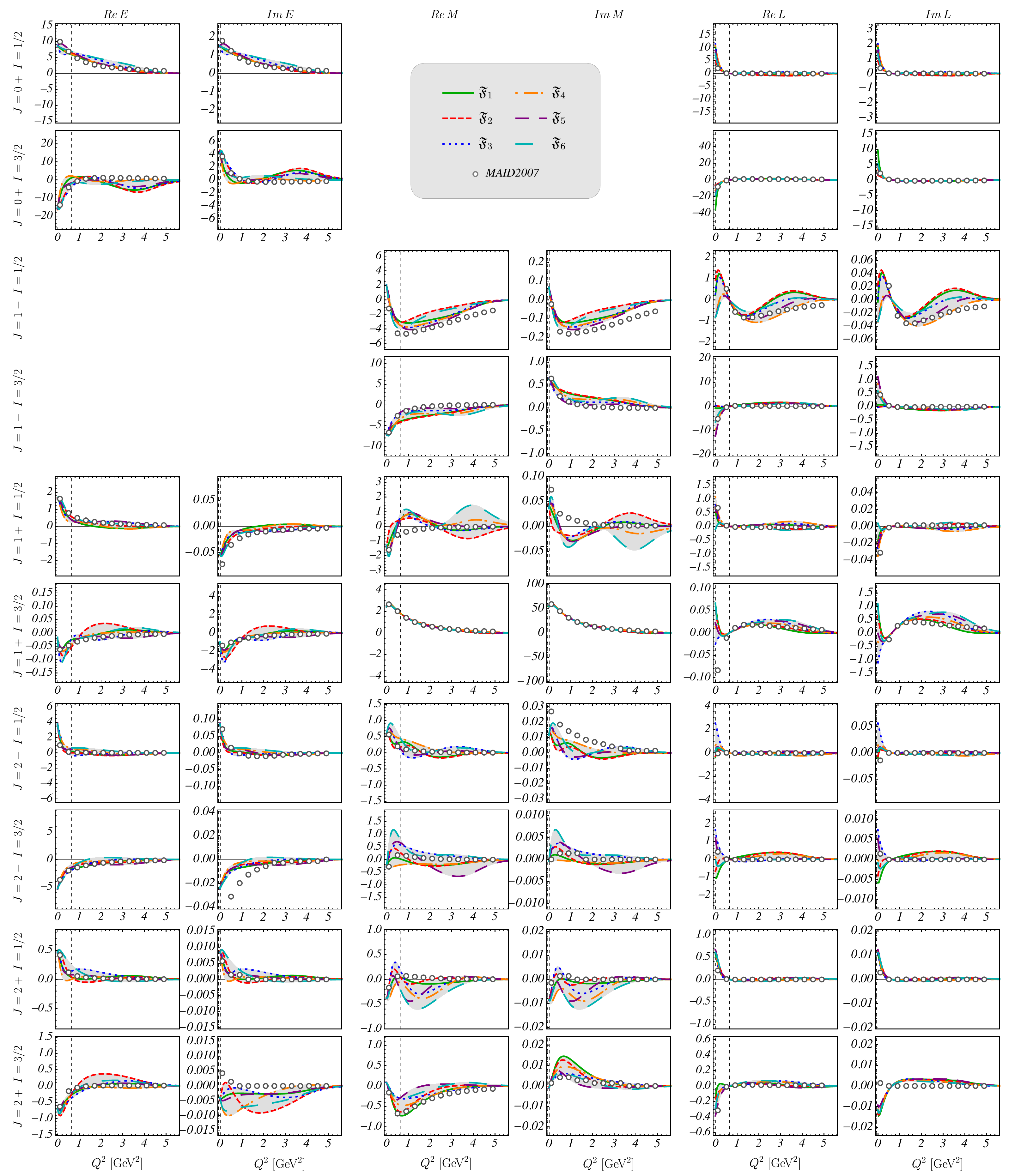}
\caption{
Fit results for multipoles in $[\rm{mfm}]$ at $W=1.230~{\rm GeV}$ as a function of $Q^2$. Different curves correspond to various fit strategies, representing systematic uncertainty of our approach -- shading between the curves is included to guide the eye. Results of MAID2007 analysis~\cite{Drechsel:2007if} are shown by open circles for comparison. Dashed vertical lines show virtualities corresponding to $q=0$ and $\omega=0$, respectively.
}
\label{fig-app:ELM-W=1230MeV}
\end{figure*}

\begin{figure*}[h!]
    \includegraphics[width=\linewidth]{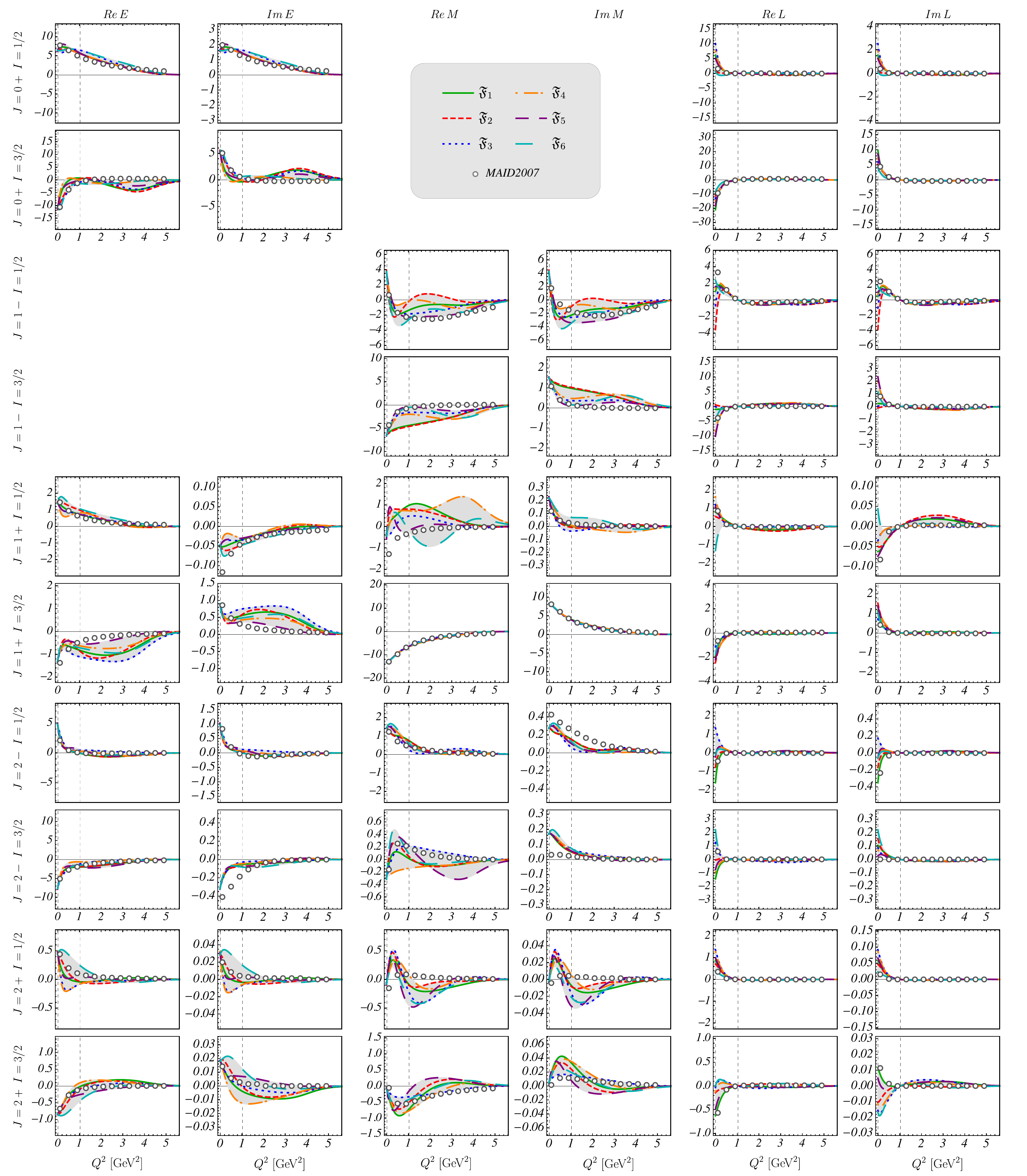}
\caption{
Fit results for multipoles in $[\rm{mfm}]$ at $W=1.380~{\rm GeV}$ as a function of $Q^2$. Different curves correspond to various fit strategies, representing systematic uncertainty of our approach -- shading between the curves is included to guide the eye. Results of MAID2007 analysis~\cite{Drechsel:2007if} are shown by open circles for comparison. Dashed vertical lines show virtualities corresponding to $q=0$ and $\omega=0$, respectively.
}
\label{fig-app:ELM-W=1380MeV}
\end{figure*}

\begin{figure*}[h!]
    \includegraphics[width=\linewidth]{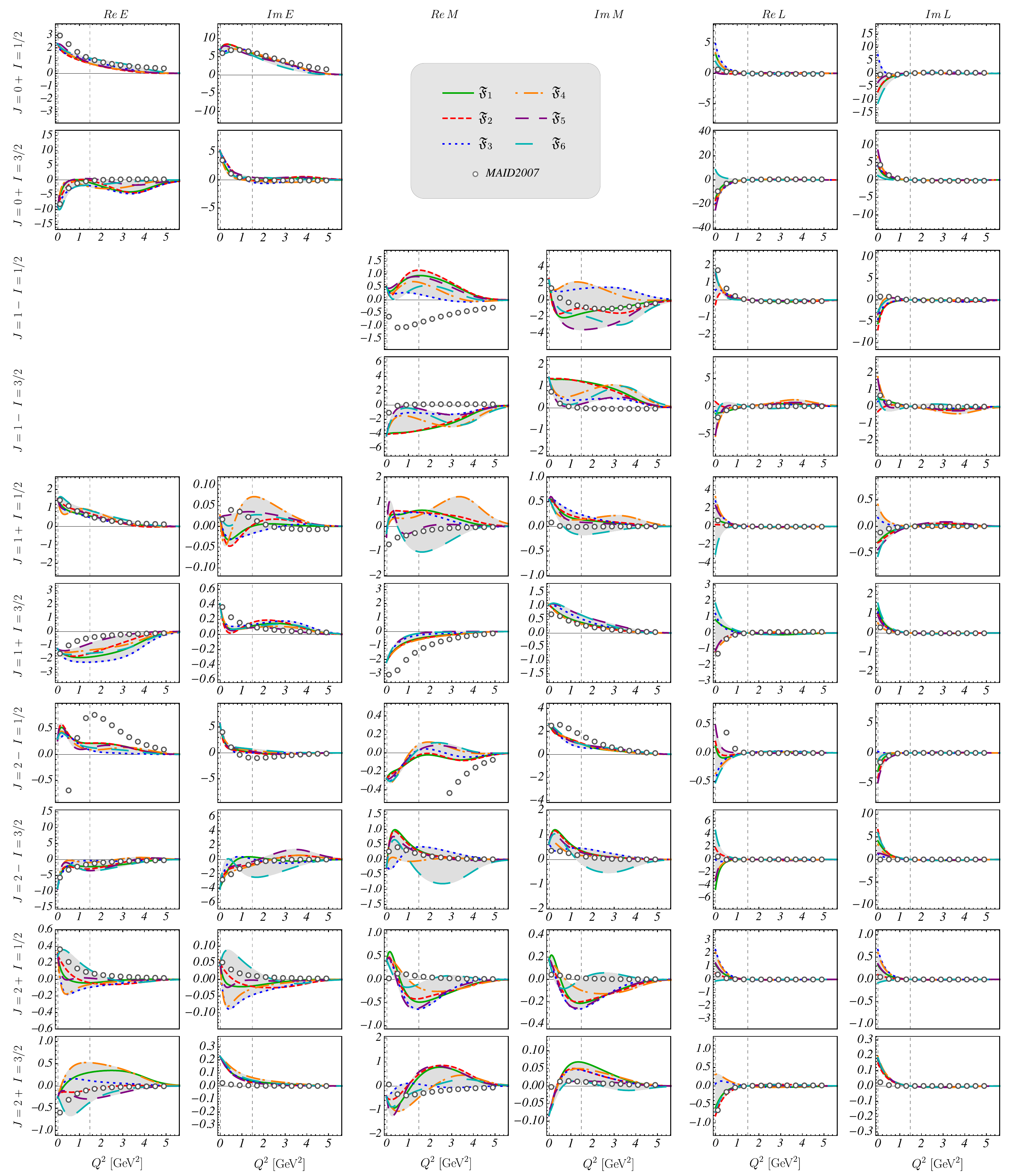}
\caption{
Fit results for multipoles in $[\rm{mfm}]$ at $W=1.535~{\rm GeV}$ as a function of $Q^2$. Different curves correspond to various fit strategies, representing systematic uncertainty of our approach -- shading between the curves is included to guide the eye. Results of MAID2007 analysis~\cite{Drechsel:2007if} are shown by open circles for comparison. Dashed vertical lines show virtualities corresponding to $q=0$ and $\omega=0$, respectively.
}
\label{fig-app:ELM-W=1535MeV}
\end{figure*}

\clearpage
\subsection{Multipoles for fixed $Q^2$}
\label{app:multipoles-Q2-fixed}

\begin{figure*}[h!]
    \includegraphics[width=\linewidth]{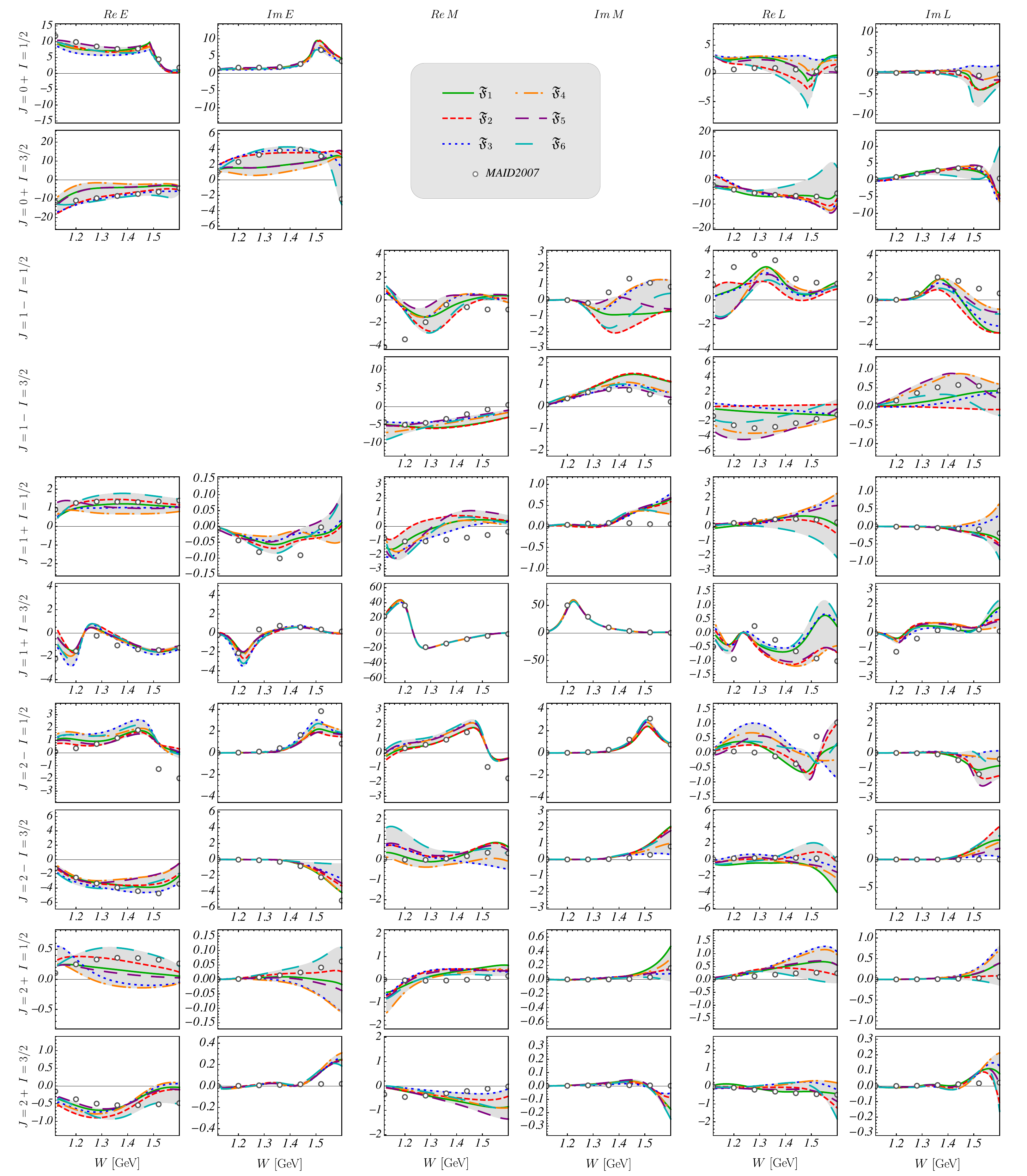}
\caption{
Fit results for multipoles in $[\rm{mfm}]$ at $Q^2=0.2~{\rm GeV}^2$ as a function of $W$. Different curves correspond to various fit strategies, representing systematic uncertainty of our approach -- shading between the curves is included to guide the eye. Results of MAID2007 analysis~\cite{Drechsel:2007if} are shown by open circles for comparison.
}
\label{fig-app:ELM-Q2=0.2}
\end{figure*}

\begin{figure*}[h!]
    \includegraphics[width=\linewidth]{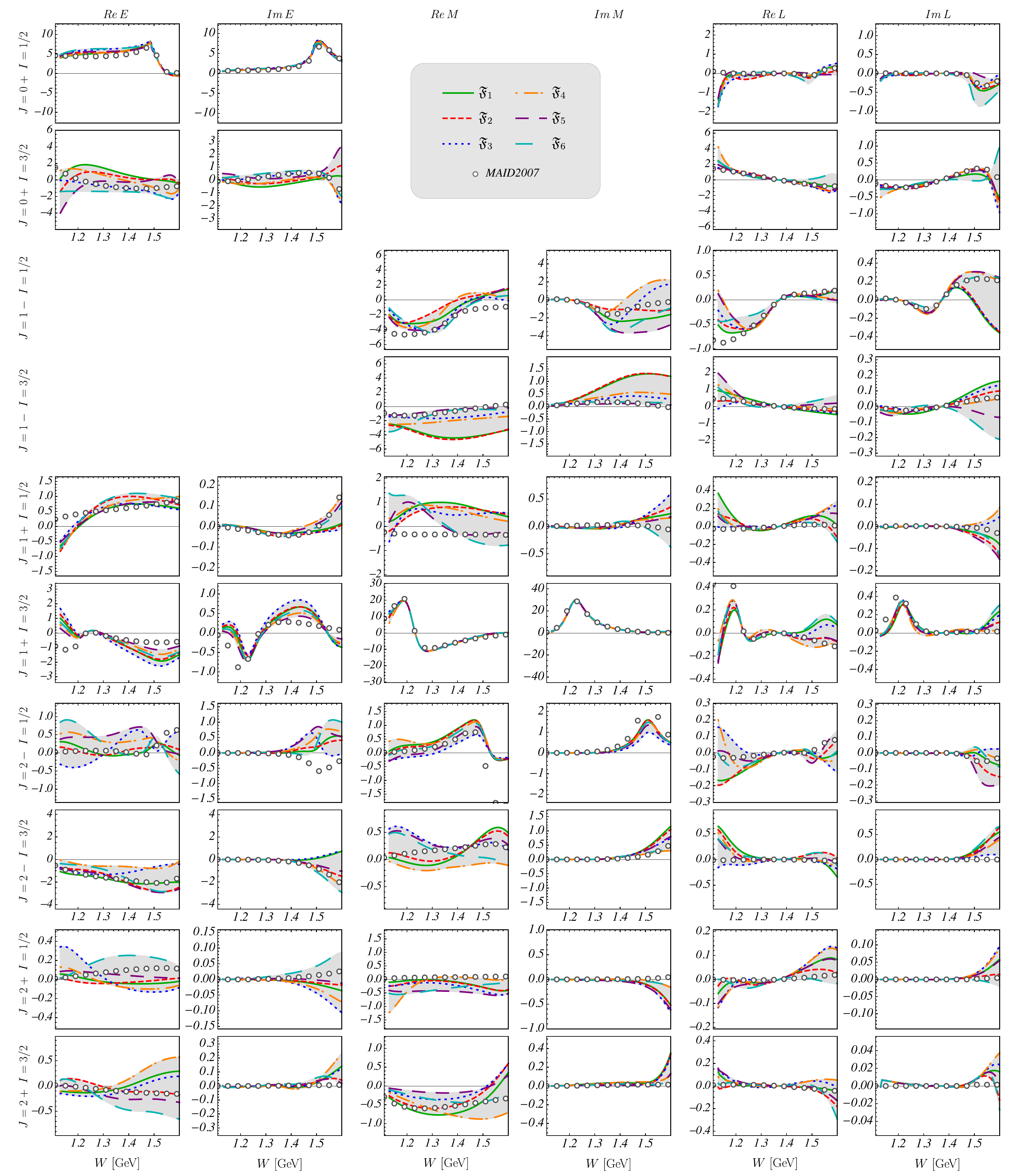}
\caption{
Fit results for multipoles in $[\rm{mfm}]$ at $Q^2=1.0~{\rm GeV}^2$ as a function of $W$. Different curves correspond to various fit strategies, representing systematic uncertainty of our approach -- shading between the curves is included to guide the eye. Results of MAID2007 analysis~\cite{Drechsel:2007if} are shown by open circles for comparison.
}
\label{fig-app:ELM-Q2=1}
\end{figure*}

\end{onecolumngrid}
\end{document}